\documentclass[useAMS,usenatbib]{mn2e}

\usepackage{graphicx}
\usepackage{bm}
\usepackage{natbib}
\usepackage{amsmath}
\usepackage{color}
\usepackage{psfrag}

\newcommand{\be}{\begin{equation}}
\newcommand{\ee}{\end{equation}}

\newcommand{\ba}{\begin{eqnarray}}
\newcommand{\ea}{\end{eqnarray}}

\newbox\grsign \setbox\grsign=\hbox{$>$}
\newdimen\grdimen \grdimen=\ht\grsign
\newbox\laxbox \newbox\gaxbox
\setbox\gaxbox=\hbox{\raise.5ex\hbox{$>$}\llap
     {\lower.5ex\hbox{$\sim$}}}\ht1=\grdimen\dp1=0pt
\setbox\laxbox=\hbox{\raise.5ex\hbox{$<$}\llap
     {\lower.5ex\hbox{$\sim$}}}\ht2=\grdimen\dp2=0pt

\def\lesssim{\mathrel{\hbox{\rlap{\hbox{\lower4pt\hbox{$\sim$}}}\hbox{$<$}}}}
\def\gtrsim{\mathrel{\hbox{\rlap{\hbox{\lower4pt\hbox{$\sim$}}}\hbox{$>$}}}}
 
\title[Magnetic Jets]{Magnetic field structure of relativistic jets without current sheets}

\author[K.~N.~Gourgouliatos et al.] {K.~N.~Gourgouliatos$^{1}$\thanks{E-mail: kgourgou@purdue.edu}, Ch.~Fendt$^2$, E.~Clausen-Brown$^1$ and M.~Lyutikov$^1$ \\
 $^1$Department of Physics, Purdue University, \\
 525 Northwestern Avenue,
West Lafayette, IN
47907-2036 \\
$^2$ Max Planck Institute for Astronomy, K\"onigstuhl 17, \\
D-69117 Heidelberg, Germany} 

\begin{document}

\date{Accepted -. Received -; in original form -}
\pagerange{\pageref{firstpage}--\pageref{lastpage}} \pubyear{-}
\maketitle

\label{firstpage}

\begin{abstract}
We present an analytical class of equilibrium solutions for the structure of relativistic sheared and rotating  magnetized jets that  contain no boundary current sheets. We demonstrate the  overall dynamical stability of  these solutions and, most importantly, a better numerical  resistive stability than the commonly employed   force-free structures which inevitably require the presence of  dissipative surface  currents.  The jet is volumetrically confined by the external pressure, with no pressure gradient on the surface.  We calculate the expected observed properties of such jets.  Given the simplicity of these solution we suggest them as  useful initial conditions for relativistic jet simulations.
\end{abstract}


\section{Introduction}
Jets are present in a great variety of astrophysical structures ranging from small scale protostellar jets \citep{Reipurth:1998} to large-scale jets feeding the enormous lobes in radio galaxies \citep{Rees:1978}. In addition, there is a broad range in jet power and lifespan, for instance $\gamma$-ray bursts release most of their energy through a jet within a few seconds \citep{Sari:1999}, whereas quasar jets are long lived entities which last for millions of years. 

Measurements of synchrotron radiation and rotation measure suggest that the presence of magnetic fields is a ubiquitous element of jets \citep{Gabuzda:2004} and it is a standard ingredient of jet models i.e. \cite{Komissarov:1999, Leismann:2005, Tchekhovskoy:2008}. The simplest approach to a magnetic jet model is that of a force-free field. In this case the magnetic field dominates the jet. So, assuming ideal MHD, the magnetic field comes to an equilibrium within a few Alfv\'en crossing times, and relaxes to a force-free state. However, any force-free magnetic field that is bound in space must have a current sheet on its boundary. A current sheet is an infinitesimal surface current which separates the non-zero component of the magnetic field which is parallel to this boundary from the external medium. Although this is a viable state for ideal MHD and it is possible to verify the stability of the field \citep{Woltjer:1958}, if one takes into account dissipative effects, surface currents are critical \citep{Taylor:1986}. Physically, surface currents shall be either sources of instability or they shall dissipate making the transition from the jet to the external medium smoother. In addition their detailed study requires a micro-physical approach upon which there is not a generally agreed picture, as in principle reconnection is described through different models \citep{Sweet:1958, Parker:1957, Petschek:1964,Uzdensky:2011}. 

A serious drawback of surface currents is their numerical treatment in simulations. A surface current formally is the derivative of a step function of the magnetic field, which is none other than a Dirac-$\delta$ function. Such discontinuities make this study a laborious task. In general, steep but finite derivatives of the magnetic field are interpreted as surface currents. Observations of distant structures cannot provide sufficient information about their presence. Solar system observations show rapid transitions in magnetic fields which are associated with large current densities, especially in structures associated to coronal mass ejections \citep{Burlaga:1981}. Nevertheless, these are explosive entities, and if the dissipation timescale is longer than their dynamical evolution they can retain surface currents. 

The elimination of surface currents is feasible through the inclusion of plasma pressure in the dynamics of the problem. In particular, when some gas pressure is included in the system the basic force-free equation is substituted by the Grad-Shafranov equation \citep{Shafranov:1966}. This equation takes into account both the force of the magnetic field on the current and that arising from the pressure gradient; the equilibrium state is given by the balance of these forces. Unlike the force-free description of the problem, this extension allows a smooth transition from the magnetized area to the external medium which contains no magnetic field. This is consistent with the result of resistive decay, which turns magnetic field energy into heat, and thus, to an increase in pressure. 

It is also possible to study the relativistic generalisation of force-free systems. Such studies have applications in the context of $\gamma$-ray bursts, AGN jets, and microquasar jets. In relativistic structures the force of the electric fields on the charges are significant and should be taken into account. Both analytical \citep{Prendergast:2005, Lyutikov:2005, Gourgouliatos:2008} and numerical progress has been made in this direction, in special and general relativistic formalism \citep{Komissarov:2002, Gammie:2003, Fendt:2004, McKinney:2006a, McKinney:2006b, McKinney:2007}. Another field of application of force-free relativistic magnetic fields is that of pulsar magnetospheres \citep{Contopoulos:1999, Goodwin:2004, Spitkovsky:2006}. Finally, it is also possible to include pressure in such systems and take into account those dynamical effects which shall give the relativistic analogue of the Grad-Shafranov equation \citep{GV:2010}.
 
In the context of relativistic jet models for launching, acceleration, and collimation, for years progress was made by relying on the assumption of steady state self-similar flows, e.g. \cite{Li:1992, Contopoulos:1995} or force-free fields \citep{Fendt:1997}.  However, recent relativistic MHD jet formation simulations have been capable of following the jet from a Keplerian disk to several thousand Schwarzschild radii above the disk \citep{Porth:2010, Porth:2011}.  Simulations in GRMHD resolve the accretion process towards the black hole, outflows, or the efficiency of a Blandford-Znajek tower, e.g. \cite{DeVilliers:2005, McKinney:2007}. 

Unlike the above work which includes jet production; we focus on the asymptotic region of the jet and work in the context of magnetic towers \citep{Lynden-Bell:2003, Uzdensky:2006} that do not have surface currents and where the field is confined inside a cylinder. Thus, our solution differs from the approach of \cite{Appl:1993} who first presented a solution to the force-free asymptotic Grad-Shafranov equation that extends across the light cylinder, but does not include return currents.   As opposed to the magnetic tower case, in our work the field is not force-free but it coexists with a plasma whose pressure has some dynamical contribution. This allows us to make the transition smoother from the area dominated by the magnetic field to the area dominated by the pressure. \cite{GBL:2010} have found a similar class of solutions for fields of topology similar to a spheromak both for static and expanding structures \citep{Lyutikov:2011, Gourgouliatos:2011}. The solutions we have found satisfy the Grad-Shafranov equation with the additional constraint that there are not surface currents on the boundary. We apply this idea to jets by solving the Grad-Shafranov equation in cylindrical geometry. We remark that the non-linearity of the Grad-Shafranov and the free parameters allow many possible equilibrium solutions, even if we assume some symmetry. A possible approach is to assume that the detailed structure of the field is dictated by the behaviour of the source of the jet. This may be true in the region very close to the origin, but we expect that further out the jet will relax to a state where quantities such as the flux and the total helicity will be conserved without memory of the fine details of the jet launching region \citep{Spruit:2010}. The problem at this stage is underdetermined, as we can construct a great number of solutions, since in principle we can choose a form for two of the three basic physical quantities appearing in the problem, i.e. poloidal and toroidal field, and solve for the third one. However any solution of this kind is not necessarily a physical configuration. A forward method of approaching the problem is the following. Consider a physical system of a spinning disc on which the field is anchored and a pressure environment. The relation between $B_{z}$ and $B_{\phi}$ shall be dictated by the way the disc spins. Because of dissipation, fields without current sheets are more likely to occur provided there is sufficient time for the field to relax in a dissipative-structure. This demand leads to a system where both components of the field go to zero on the boundary. Then we can determine the pressure inside the jet through the Grad-Shafranov equation, but it also has to be such that the total energy carried by the system is a minimum for the given boundary conditions, so that the equilibrium is stable. Therefore the physical problem is summarized in the following steps. Some poloidal flux is connected to a disc, which spins and generates some toroidal flux. Inside the jet there is some pressure, which is lower in regions  with a stronger magnetic field, and increases as one moves to regions of weaker magnetic field and dominates totally outside of the jet. The family of solutions of the Grad-Shafranov equation gives combinations of functions which correspond to equilibria, but not necessarily to the most economic ones in terms of energy. So we demand that the energy is also minimum for given boundary conditions.  Although the above process is natural, it is far from being analytically soluble. Using the experience of force-free fields where stable solutions are given for fields where the same amount of current is carried by each field line, which are none others than the constant $\alpha$ solutions \citep{Tayler:1973}, we shall relate linearly the pressure to the poloidal or the toroidal flux. Then, by performing some preliminary simulations we shall verify that the fields are not destroyed by instabilities. 

We consider these solutions useful for two reasons. First they may be realistic physical states after dynamical and dissipative relaxation has taken place, the dynamical relaxation leads to force equilibrium, while the dissipative relaxation leads to the elimination of surface currents and a possible scenario is that of mixing of the gas with the magnetic field. When the timescales of the jet are shorter than the ones of the external medium the system has time to relax into a stable equilibrium. This is true for a jet with lower density compared to the external medium, as the Alfv\'en velocity, determining the timescale of the jet is faster than the sound speed of the denser external medium. Second we suggest that these magnetic structures can be used as trial solutions or limiting states in jet simulations. They are simple enough so that they can be used without significant modification in the existing simulations. In addition the simulations we have performed demonstrate that indeed, they are viable models.

We shall present two types of systems. The first one corresponds to the static problem where there is no motion at all. Such configurations have only magnetic field and gas pressure which are in equilibrium. The second one is the treatment of relativistic outflows. The plasma moves parallel to the axis, but the time derivatives of the physical quantities are zero, giving stationary solutions. This configuration contains both electric and magnetic fields and plasma pressure. The electric field is induced by motion parallel to the axis of the cylinder. Therefore the forces that come to equilibrium are the force of the magnetic field on the electric current, the force of the electric field on the charges and the gradient of the gas pressure.

\section{Static jets}

\begin{figure}
	\centering
		\includegraphics[width=0.45\textwidth]{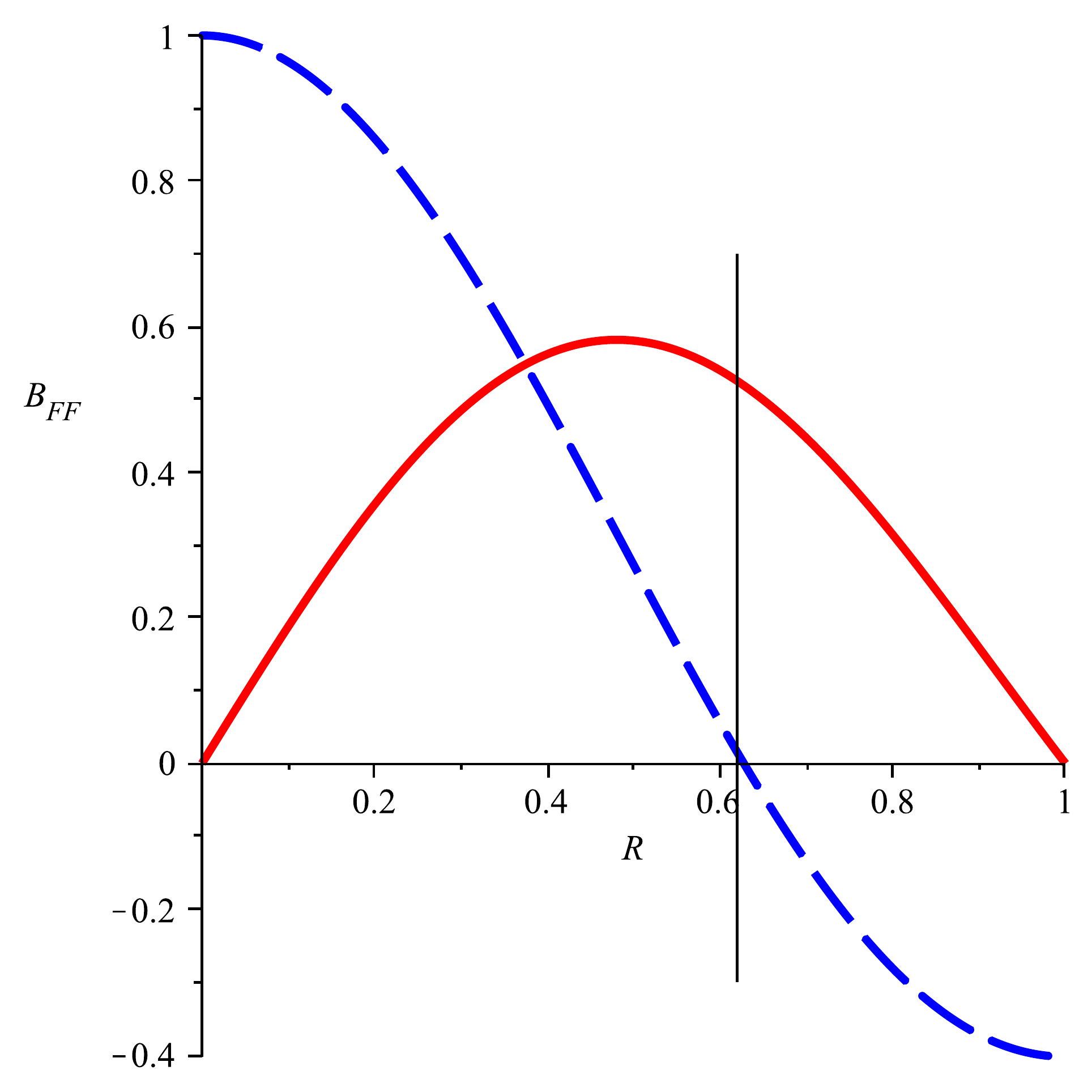}
		\caption{The Lundquist force-free magnetic field. The solid line is the $B_{\phi}$ component of the field and the dashed is the $B_{z}$. There is not a natural end of the confined field, one can choose to end the jet where the $B_{\phi}$ field becomes zero for the first time $R=1$, in this case the $B_{z}$ component has already reversed its polarity, or earlier where at $R=0.63$ where the $B_{z}$ becomes zero for the first time (vertical line). The fields never become zero simultaneously thus there is no bound magnetic field configuration without surface currents.}
		\label{FigFF}
\end{figure}

In this section we set the mathematical framework of the non-relativistic static problem. The basic equation we solve is the Grad-Shafranov equation $\frac{1}{c}\bm{j}\times \bm{B}=\nabla p$, the current is given by $\nabla \times \bm{B}=\frac{4 \pi}{c}\bm{j}$. 
\begin{eqnarray}
(\nabla \times \bm{B}) \times \bm{B}=4 \pi \nabla p\,,
\label{GS}
\end{eqnarray}
where $\bm{B}$ is the magnetic field and $p$ is the plasma pressure. The Grad-Shafranov differential equation is a non-linear partial differential equation and its solution is a rather complicated task. In general some assumption of symmetry is made, and the physical quantities appearing can be expressed in terms of two coordinates, but still the field can have components in all three dimensions. The physical quantities appearing are the poloidal field, the toroidal field and the pressure. In this paper we examine the most fundamental case, where the field is cylindrically symmetric, namely it is symmetric under translations and rotations with respect to an axis, therefore the physical quantities shall depend only in the radial coordinate. We shall express the field in two different ways. The field written in terms of the poloidal flux $P_{p}$ and poloidal current $I_{p}$ is
\begin{eqnarray}
\bm{B}=\nabla P_{p}(R) \times \nabla \phi +I_{p}(R) \nabla \phi\,,
\end{eqnarray}
and we can also write it in terms of $P_{t}$ which is related to the toroidal flux: $\int B_{\phi}dR=P_{t}(R)+$const., by an additive constant, note that as the system is symmetric under translations in $z$ there is no need to integrate along this coordinate. 
\begin{eqnarray}
\bm{B}=\nabla P_{t}(R) \times \nabla z +I_{t}(R)\nabla z\,.
\end{eqnarray}
In the absence of pressure the field comes to a force-free equlibrium which is given by $\nabla \times \bm{B}=\alpha\bm{B}$. The most stable fields are given for a spatially constant $\alpha$ \citep{Tayler:1973}. When we write the fields in terms of the fluxes we find that the constant $\alpha$ solutions correspond to fields for which $I_{p,t}=\alpha_{p,t} P_{p,t}$. The final solution, independent of the choice of the representation is that of the \cite{Lundquist:1951} field:
\begin{eqnarray}
\bm{B}_{FF}=c_{0}\alpha (J_{1}(\alpha R) \hat{\bm{\phi}} + J_{0}(\alpha R)\hat{\bm{z}})\,,
\end{eqnarray}
where $c_{0}$ is a normalisation constant and $J_{0}$ and $J_{1}$ are the Bessel functions of zeroth and first order, Fig.~(\ref{FigFF}).

\begin{figure}
	\centering
		\includegraphics[width=0.45\textwidth]{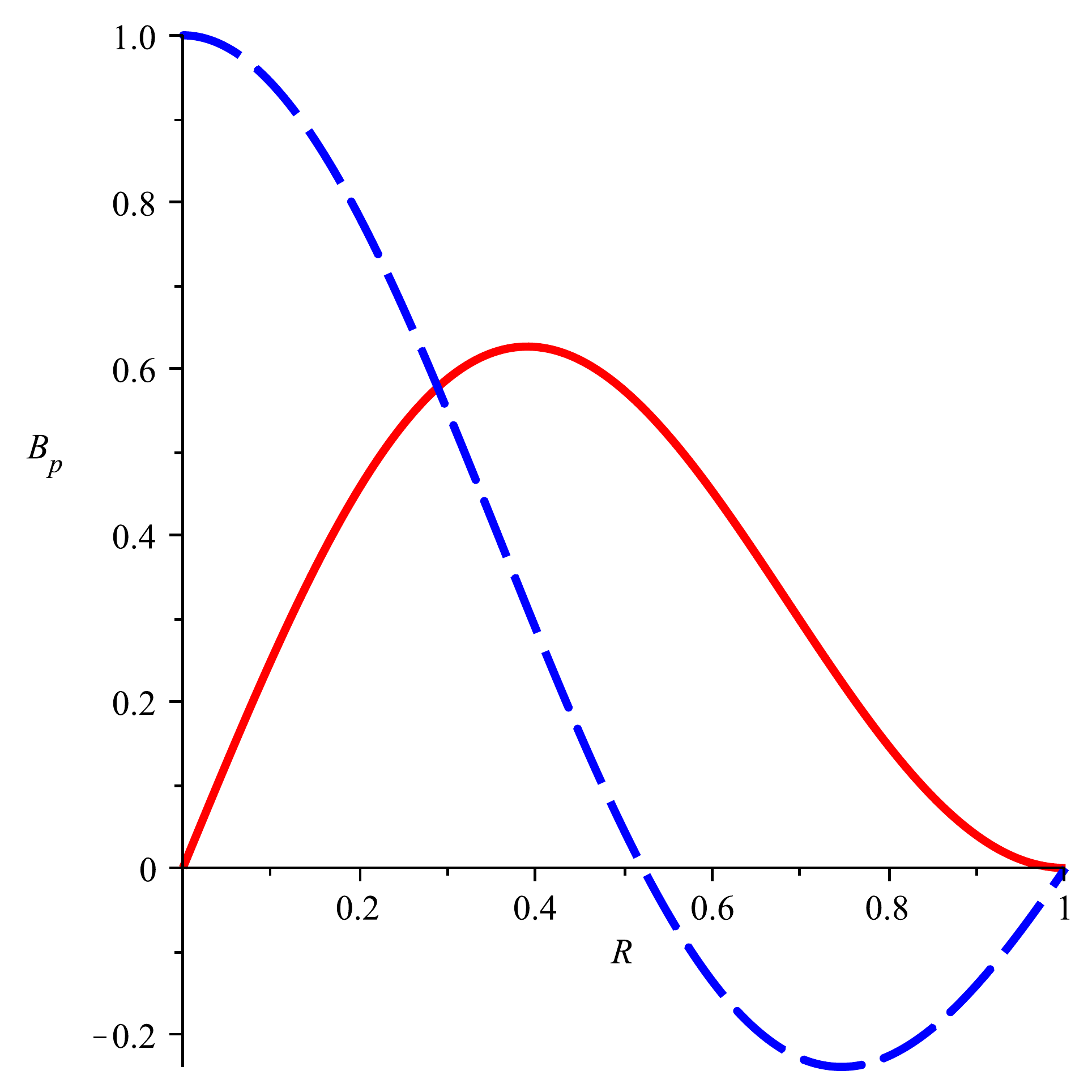}
		\caption{The magnetic field arising from $P_{p}$, the solid line is the $B_{\phi}$ component of the field and the dashed is the $B_{z}$. Note that $B_{z}$ changes polarity before $B_{\phi}$ reaches zero, resembling the behaviour of the force-free Lundquist field.}
		\label{Fig1}
\end{figure}

\begin{figure}
	\centering
		\includegraphics[width=0.45\textwidth]{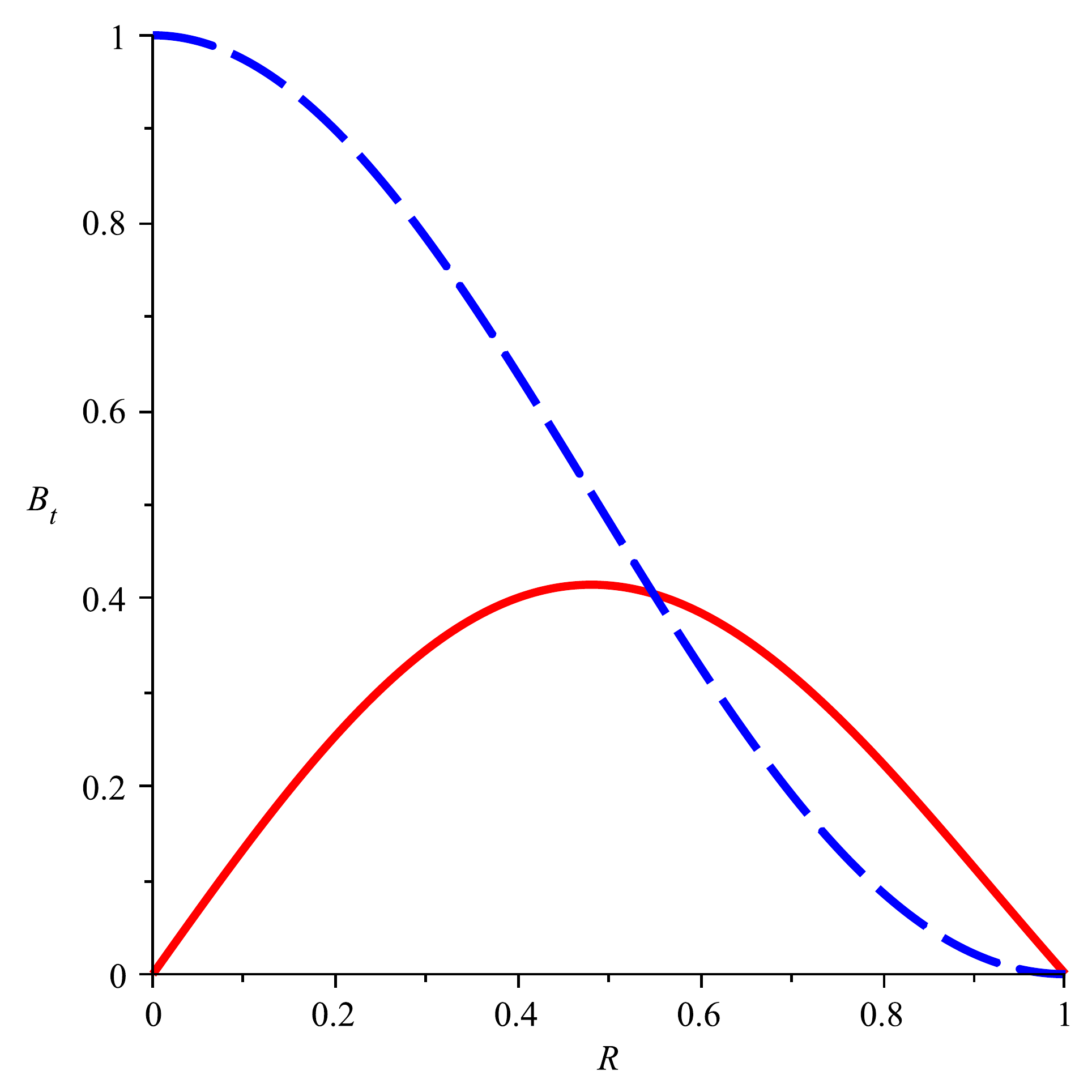}
		\caption{The magnetic field arising from $P_{t}$, the solid line is the $B_{\phi}$ component of the field and the dashed is the $B_{z}$. Note that these fields correspond to physically distinguishable structures from the case presented in Fig.~(\ref{Fig1}) and the force-free Lundquist field, Fig.~(\ref{FigFF}).}
		\label{Fig2}
\end{figure}

\begin{figure}
	\centering
		\includegraphics[width=0.45\textwidth]{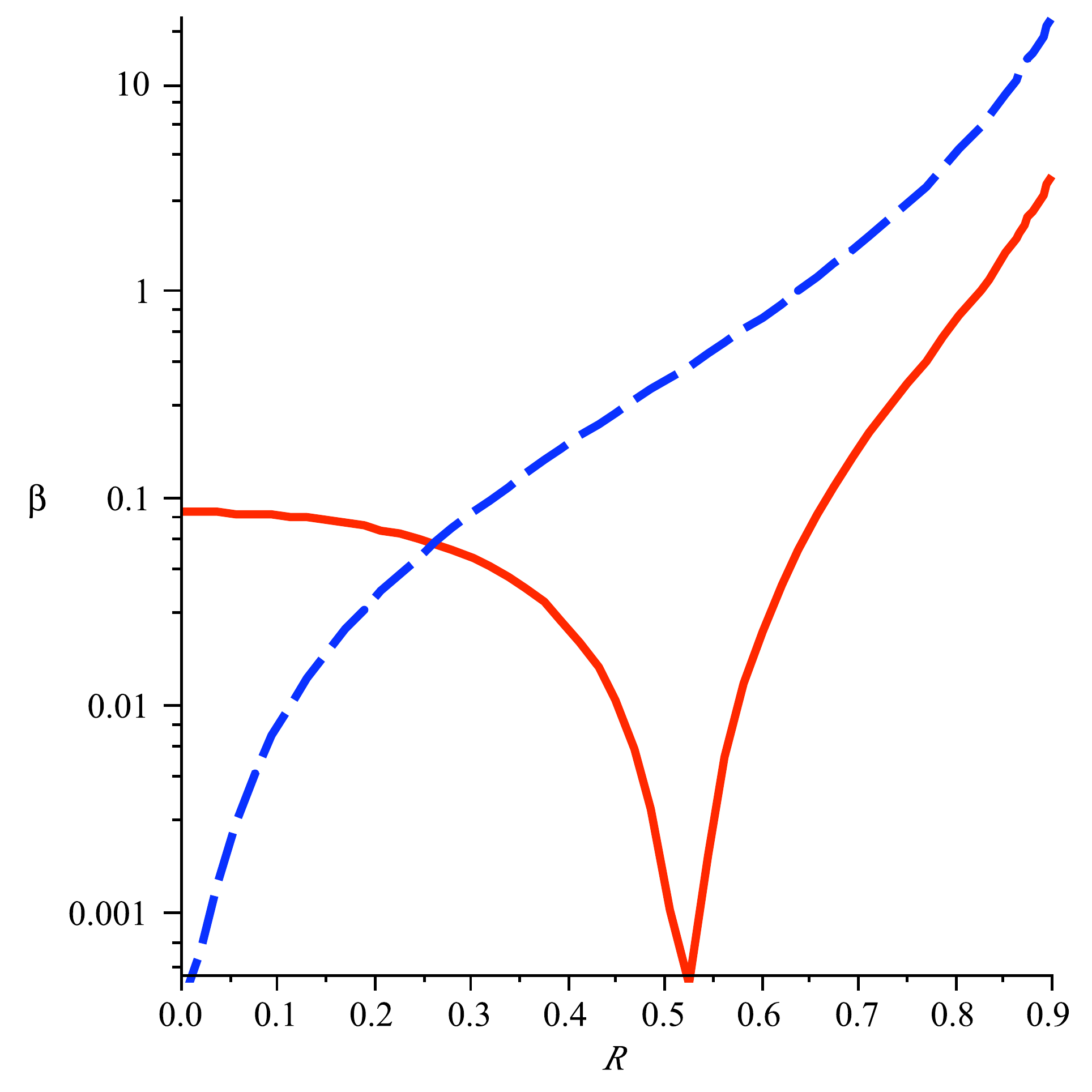}
		\caption{Logarithmic plot for the value of plasma $\beta$ for the lowest acceptable value of $p_{0~p,t}$, as a function of $R$ in units of $R_{\rm jet}$. The solid red curve corresponds to the poloidal solution and the blue dashed curve to the toroidal solution. The value of $\beta$ becomes zero at $0.52~R_{\rm jet}$ for the poloidal-type solution; and on the axis for the toroidal-type solution as the plasma pressure is zero there.  In contrast, it becomes infinite as $R\rightarrow R_{\rm jet} $ where the magnetic field is zero.  }
		\label{BETA}
\end{figure}

In the Grad-Shafranov equation there is an extra force arising from the pressure gradient. Analytical solutions are possible under the assumption of a pressure profile proportional to the flux. Unlike the force-free case, the solutions we find are different if we take the pressure to be associated with the poloidal or toroidal flux. So we write $p_{p,t}=\frac{1}{4\pi}F_{p,t}P_{p,t}+p_{0~p,t}$, where $F_{p,t}$ are constants, and $p_{0~p,t}$ is an additive constant, which does not have any dynamical role in the problem, but it has a lower limit to ensure that the plasma pressure is positive everywhere. Substituting those in equation~(\ref{GS}) and solving for the two cases we find that
\begin{eqnarray}
P_{p}=c_{p}RJ_{1}(\alpha_{p}R)-\frac{F_{p}R^{2}}{a_{p}^{2}}\,,
\label{SOL1}
\end{eqnarray}
\begin{eqnarray}
P_{t}=c_{t}J_{0}(\alpha_{t}R)-\frac{F_{t}}{\alpha_{t}^{2}}\,,
\label{SOL2}
\end{eqnarray}
where $c_{p,t}$ are normalisation constants. We are looking for solutions without surface currents confined to a cylinder of unit radius $R_{\rm jet}=1$, thus in both solutions we want $P_{p,t}(1)=0$ and $P_{p,t}'(1)=0$. Applying these conditions and choosing a normalisation so that the maximum value of the $B_{z}$ field is unity we find $c_{p}=0.172$, $\alpha_{p}=5.14$, $F_{p}=-1.54$, and $c_{t}=0.186$, $\alpha_{t}=3.83$, $F_{t}=-1.10$.
These two solutions give two physically distinct fields
\begin{eqnarray}
\bm{B}_{p}=(c_{p}\alpha_{p}J_{1}(\alpha_{p}R)-\frac{F_{p}R}{\alpha_{p}})\hat{\bm{\phi}}+\nonumber \\
(c_{p}\alpha_{p}J_{0}(\alpha_{p}R)-\frac{2F_{p}}{\alpha_{p}^{2}})\hat{\bm{z}}\,,
\end{eqnarray}
\begin{eqnarray}
\bm{B}_{t}=c_{t}\alpha_{t}J_{1}(\alpha_{t}R)\hat{\bm{\phi}}+(c_{t}\alpha_{t}J_{0}(\alpha_{t}R)-\frac{F_{t}}{\alpha_{t}})\hat{\bm{z}} \,,
\end{eqnarray}
which are plotted in Figs.~(\ref{Fig1}, \ref{Fig2}). The value for plasma $\beta=8 \pi p/ B^{2}$ that we take for the lowest value of $p_{0~p,t}$ is plotted in Fig.~(\ref{BETA}). For both configurations, in the greatest part of the jet $\beta<1$, including regions where $\beta<<1$ where the jet is practically force-free.
Even if we do not choose to express the field in terms of some flux we can write $\bm{B}=B_{\phi}\hat{\bm{\phi}}+B_{z}\hat{\bm{z}}$ and the Grad-Shafranov equation is written in the form
\begin{eqnarray}
\frac{R}{2}(B_{z}^{2}+B_{\phi}^{2}+8 \pi p)'+B_{\phi}^{2}=0\,.
\label{MOMNR}
\end{eqnarray}
Despite the fact that this equation does not add new information in the problem it shall give useful insight when we approach the outflow problem. 

\section{Sheared jets}

\subsection{General solution}
\label{GENERAL}

In this section we study outflows parallel to the axis. This procedure can work for any velocity, from slow ones to relativistic. To model the problem, we keep the cylindrical configuration of the static problem but we impose a velocity along the $z$-axis $\bm{v}=v_{z}\hat{\bm{z}}$, whose profile is cylindrically symmetric, independent of time, but varies with radius. This velocity induces a radial electric field, whose divergence creates an electric charge density and electric forces appear. As the magnetic field has no $R$ component the field lines are not stretched and retain their shape. In addition there is no time dependence thus we do not expect displacement currents. The magnetic field is
\begin{eqnarray}
\bm{B}=B_{\phi}(R)\hat{\bm{\phi}}+B_{z}(R)\hat{\bf{z}}\,,
\end{eqnarray}
and the electric field
\begin{eqnarray}
\bm{E}=E_{R}(R)\hat{\bm{R}}\,,
\end{eqnarray}
which is associated to the velocity by
\begin{eqnarray}
E_{R}=v_{z}B_{\phi}\,.
\label{ELECTRIC}
\end{eqnarray}
The time derivative of the magnetic field is zero and the electric field is curl-free, thus the Faraday induction equation $\nabla \times \bm{E}=-\frac{1}{c}\frac{\partial \bm{B}}{dt}=0$ holds. 
The electric charge density is 
\begin{eqnarray}
j^{0}=\frac{c}{4 \pi}\nabla \cdot \bm{E}\,,
\end{eqnarray}
the electric current density is
\begin{eqnarray}
\bm{j}=\frac{c}{4 \pi}\nabla \times \bm{B}\,,
\end{eqnarray}
and we can verify that Gauss's law for the magnetic field  $\nabla \cdot \bm{B}=0$ holds. 
The momentum equation is
\begin{eqnarray}
\Gamma \rho_{0}\Big(\frac{\partial}{\partial t}+\bm{v}\cdot\nabla\Big)(\xi \Gamma \bm{v})-\nabla p+\frac{j^{0}\bm{E}+\bm{j}\times \bm{B}}{c}=0\,,
\label{MOM}
\end{eqnarray}
where $\Gamma=(1-\frac{v^{2}}{c^{2}})^{-1/2}$ is the Lorentz factor, $\rho_{0}$ is the rest mass density and $\xi$ is the relativistic specific enthalpy over $c^{2}$. The first term of the momentum equation~(\ref{MOM}) appearing in brackets is zero as there is no time variation and the velocity is along $z$ so from the $\nabla$ operator we shall keep only the derivative with respect to $z$, but since our system does not have a $z$ dependence its contribution is zero. We remark that if we chose another velocity profile which contained a $\phi$ component, inertial forces should appear. This case is examined in greater detail in the simulation part of the paper. In the analytic part we focus on the axial motion, where the equilibrium is reached through the balance of field forces and pressure. For the present we do not need to take into account the details of the mass density and the relativistic specific enthalpy of the system provided that they do not vary with time or $z$. Thus we are left with the last two terms involving the pressure and the electromagnetic fields. Now we proceed to the solution of equation~(\ref{MOM}) which after substituting the fields can be written as
\begin{eqnarray}
\frac{R}{2}(B_{z}^{2}+B_{\phi}^{2}-E_{R}^{2}+8 \pi p)'+B_{\phi}^{2}-E_{R}^{2}=0\,.
\label{MOMREL}
\end{eqnarray}
Note the similarity with equation (\ref{MOMNR}), with the extra term $-E_{R}^{2}$ which is a result of the relativistic invariance of $B^{2}-E^{2}$. For that reason, we shall introduce the new variable
\begin{eqnarray}
H^{2}=B_{\phi}^{2}-E_{R}^{2}\,,
\label{DEFH}
\end{eqnarray}
which can be substituted to equation~(\ref{MOMREL}) and it shall give
\begin{eqnarray}
\frac{R}{2}(B_{z}^{2}+H^{2}+8 \pi p)'+H^{2}=0\,.
\end{eqnarray}
Based on the results of the previous section, we can generalise the solutions we have found in the previous section. In the static regime we had two representations corresponding to the flux functions $P_{p}$ and $P_{t}$. Here instead we shall use the generalised functions $G_{p}$ and $G_{t}$. The first representation is 
\begin{eqnarray}
H=\frac{\alpha_{p}G_{p}}{R}\,, \nonumber \\
B_{z}=\frac{1}{R}\frac{dG_{p}}{dR}\,, \nonumber \\
p=\frac{1}{4\pi}F_{p}G_{p}+p_{p,0}\,.
\end{eqnarray}
The second representation is
\begin{eqnarray}
H=-\frac{dG_{t}}{dR}\,, \nonumber \\
B_{z}=\alpha_{t}G_{t}\,, \nonumber \\
p=\frac{1}{4\pi}F_{t}G_{t}+p_{t,0}\,.
\end{eqnarray}
The solution to the differential equation is exactly the same as the one in the static problem and is given by equations (\ref{SOL1}) and (\ref{SOL2}), while the values of $F_{p,t}$ and $\alpha_{p,t}$ are the ones discussed there, for the same boundary conditions, namely the boundary of the jet is at $R_{\rm jet}=1$ and does not have surface currents. Then we can evaluate the electric field and the azimuthal component of the magnetic field, from equations (\ref{ELECTRIC}) and (\ref{DEFH}) 
\begin{eqnarray}
B_{\phi}^{2}=\frac{H^{2}}{1-v^{2}_{z}}\,,
\end{eqnarray}
\begin{eqnarray}
E_{R}^{2}=\frac{v^{2}_{z}}{1-v^{2}_{z}}H^{2}
\end{eqnarray}
Now the issue is the velocity profile we choose which gives the partition of $H$ in the electric and magnetic field.

\subsection{Solutions for a given velocity profile}

Having solved the problem for the generic velocity profile we can write the expressions for the fields. The field and the pressure for the poloidal-flux approach are: 
\begin{eqnarray}
E_{R}=\frac{v_{z}}{(1-v_{z}^{2})^{1/2}}\Big(c_{p}\alpha_{p}J_{1}(\alpha_{p}R)-\frac{F_{p}R}{\alpha_{p}}\Big)\,,
\end{eqnarray}
\begin{eqnarray}
B_{\phi}=\frac{1}{(1-v_{z}^{2})^{1/2}}\Big(c_{p}\alpha_{p}J_{1}(\alpha_{p}R)-\frac{F_{p}R}{\alpha_{p}}\Big)\,,
\end{eqnarray}
\begin{eqnarray}
B_{z}=c_{p}\alpha_{p}J_{0}(\alpha_{p}R)-\frac{2F_{p}}{\alpha^{2}_{p}}\,,
\end{eqnarray}
\begin{eqnarray}
p=\frac{1}{4\pi}F_{p}\Big(c_{p}RJ_{1}(\alpha_{p}R)-\frac{F_{p}R^{2}}{\alpha^{2}_{p}}\Big)+p_{p,0}\,,
\end{eqnarray} 
\begin{figure}
	\centering
		\includegraphics[width=0.45\textwidth]{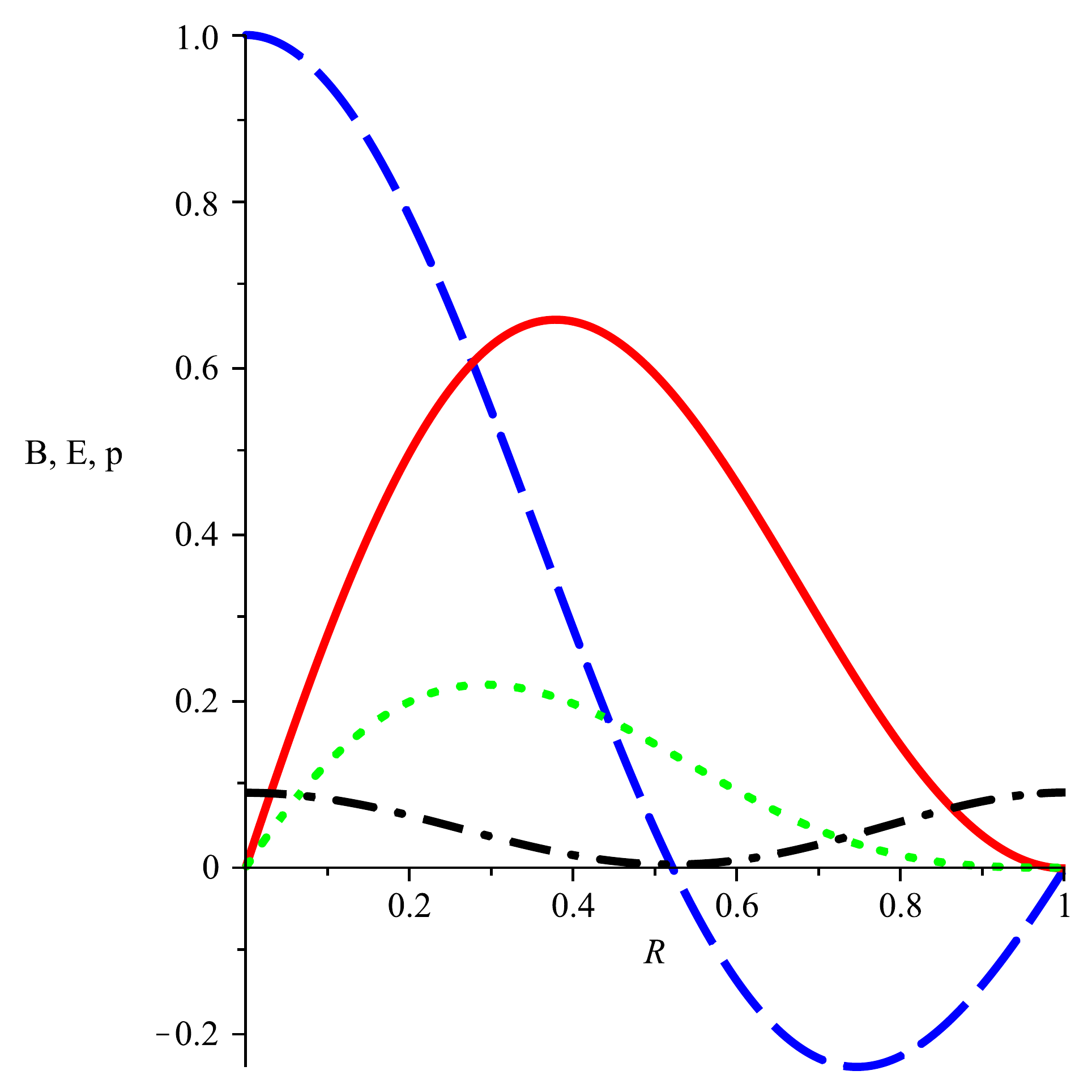}
		\caption{The magnetic and electric fields and the pressure arising from $G_{p}$. The solid line is the $B_{\phi}$ component of the field, the dashed is the $B_{z}$, the dotted the electric field and the dotted-dashed line is the $p$, on which a suitable value for $p_{p,0}$ has been chosen so that it is positive, note that the fields and the pressure are of different dimensions so there shall be no comparison in the plot.}
		\label{REL1}
\end{figure}
and for the toroidal-flux approach are:
\begin{eqnarray}
E_{R}=\frac{v_{z}}{(1-v_{z}^{2})^{1/2}}c_{t}\alpha_{t}J_{1}(\alpha_{t}R)\,,
\end{eqnarray}
\begin{eqnarray}
B_{\phi}=\frac{1}{(1-v_{z}^{2})^{1/2}}c_{t}\alpha_{t}J_{1}(\alpha_{t}R)\,,
\end{eqnarray}
\begin{eqnarray}
B_{z}=\alpha_{t}\Big(c_{t}J_{0}(\alpha_{t}R)-\frac{F_{t}}{\alpha_{t}^{2}}\Big)\,,
\end{eqnarray}
\begin{eqnarray}
p=\frac{1}{4\pi}F_{t}\Big(c_{t}J_{0}(\alpha_{t}R)-\frac{F_{t}}{\alpha^{2}_{t}}\Big)+p_{t,0}\,,
\end{eqnarray}
\begin{figure}
	\centering
		\includegraphics[width=0.45\textwidth]{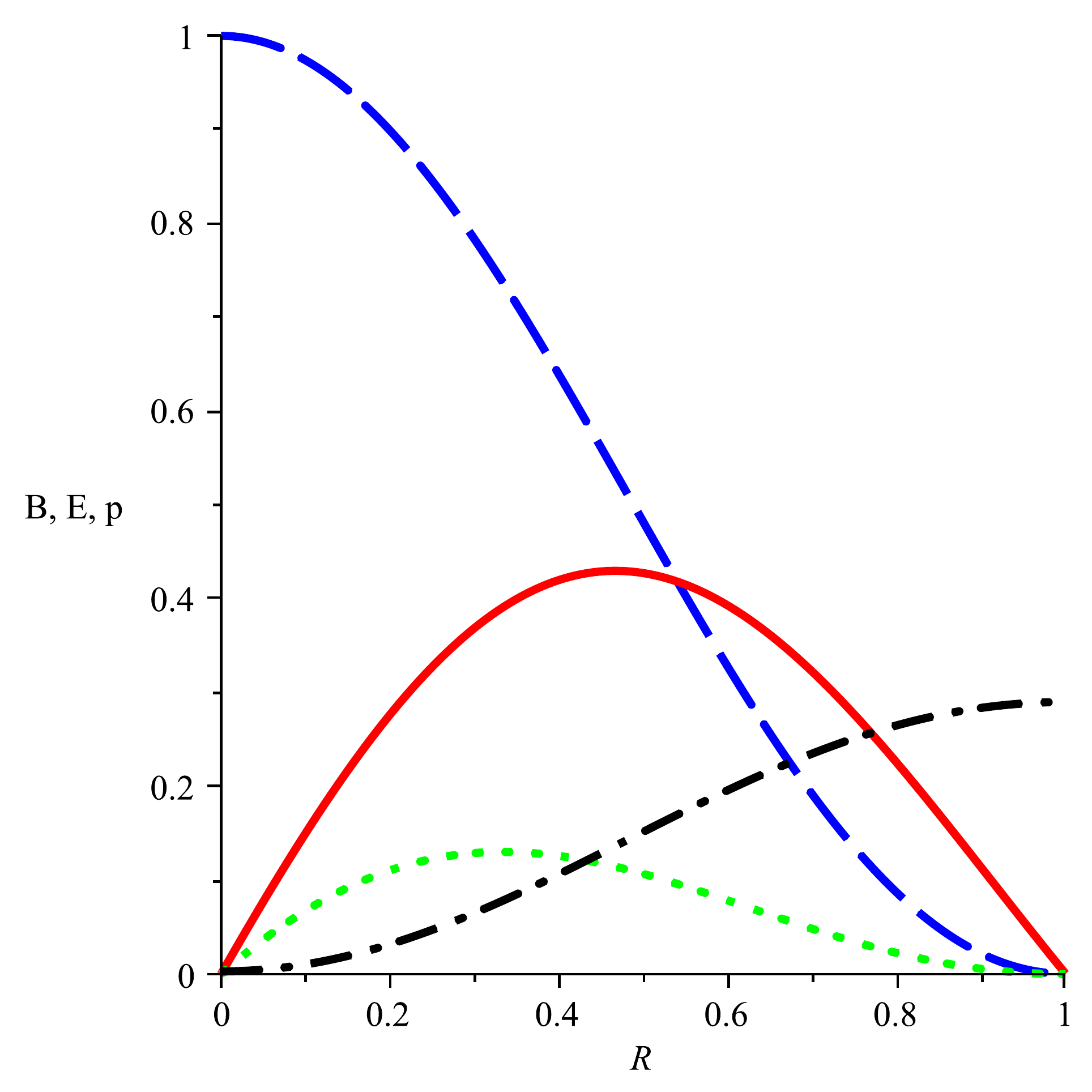}
		\caption{The magnetic and electric fields and the pressure arising from $G_{t}$. The solid line is the $B_{\phi}$ component of the field, the dashed line is the $B_{z}$, the dotted line is the electric field and the dotted-dashed line is the $p$, on which a suitable value for $p_{t,0}$ has been chosen so that it is positive.  Note that the fields and the pressure are of different dimensions so there shall be no comparison in the plot.}
		\label{REL2}
\end{figure}
For instance, we choose a linear velocity profile which is maximum on the axis and zero at $R=1$, $v_{z}=v_{z,0}(1-R)$. For this particular case the fields have the forms plotted in Figs.~(\ref{REL1}) and (\ref{REL2}). Note that the values of $\alpha_{p,t}$ and $F_{p,t}$ are those found in the static problem, so that the jet is confined to a cylinder of unit radius with no surface currents. 

\begin{figure}
	\centering
		\includegraphics[width=0.45\textwidth]{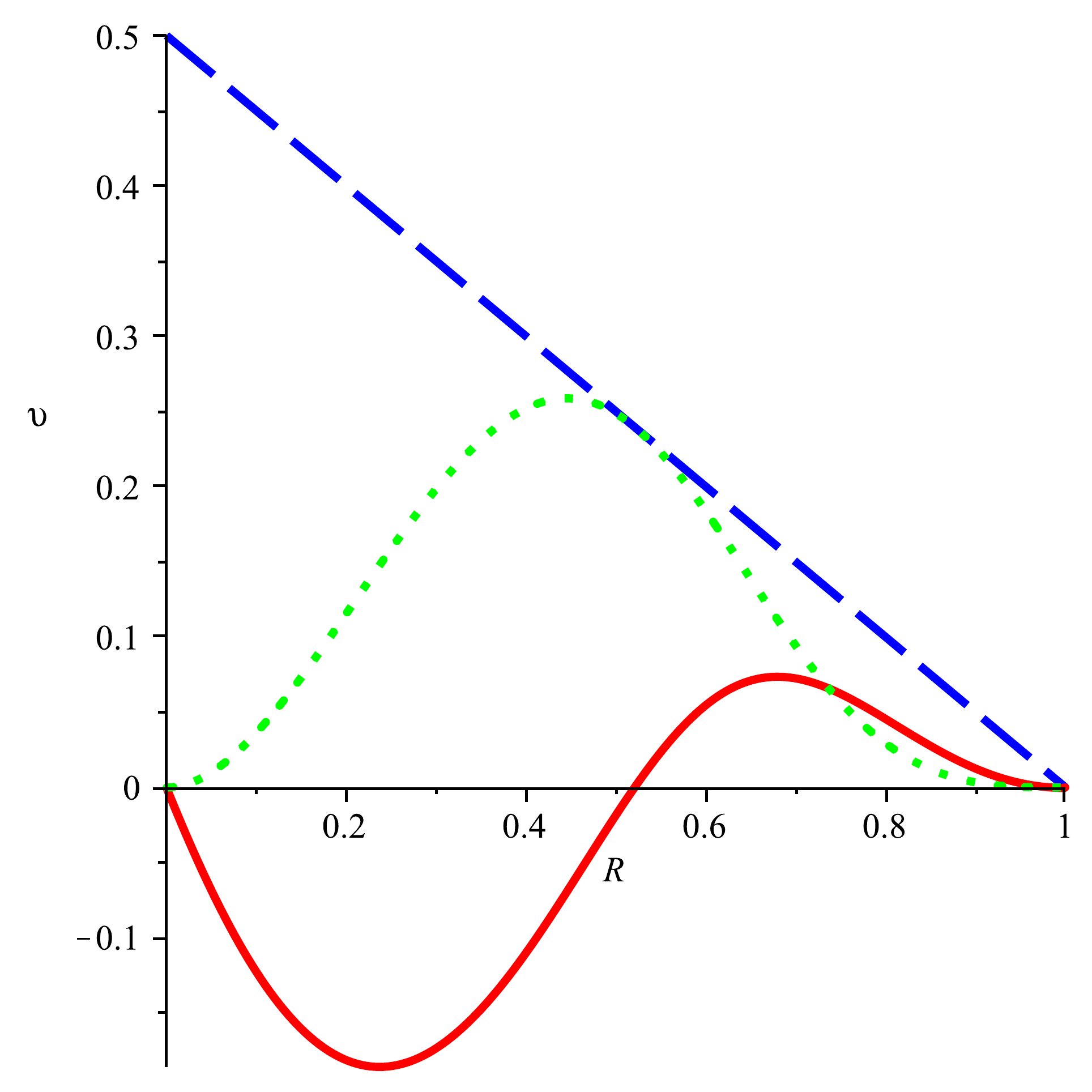}
		\caption{The velocities of the flow and the field lines normalised to $c$ for the field found in the poloidal-flux approach, Fig.~(\ref{REL1}). The solid line is the toroidal component of $\bm{V}_{F}$, the dotted line is the axial component of $\bm{V}_{F}$ and the dashed line is $\bm{v}$ which is chosen to be in the $z$ direction. In this case we have chosen that the maximum velocity for $\bm{v}$ to be $0.5c$.}
		\label{VELOCITY1}
\end{figure}

\begin{figure}
	\centering
		\includegraphics[width=0.45\textwidth]{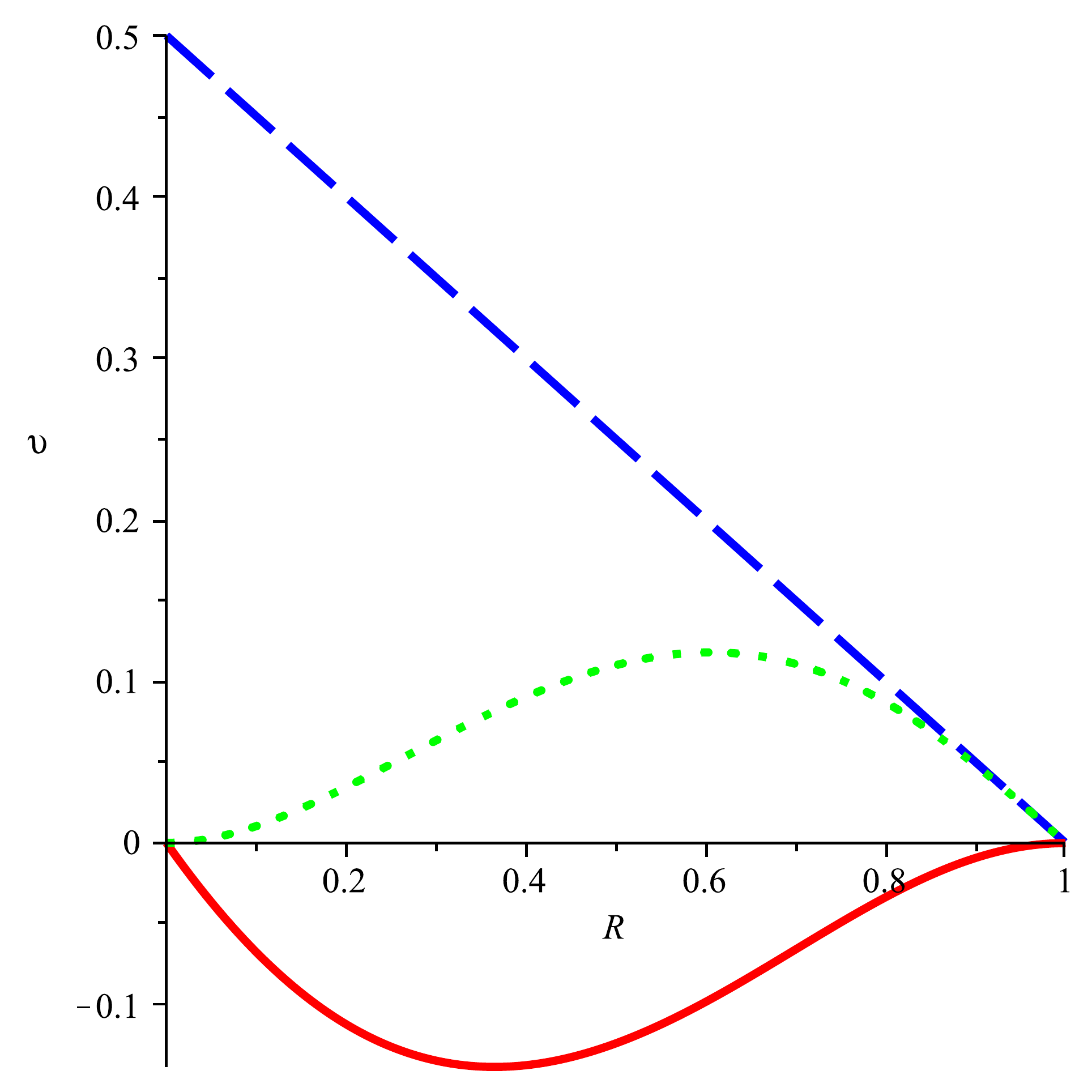}
		\caption{The velocities of the flow and the field lines normalised to $c$, for the field found in the toroidal-flux approach, Fig.~(\ref{REL2}). The solid line is the toroidal component of $\bm{V}_{F}$, the dotted line is the axial component of $\bm{V}_{F}$ and the dashed line is $\bm{v}$ which is chosen to be in the $z$ direction. Note that we use the same profile for $\bm{v}$ as in the case of Fig.~(\ref{VELOCITY1}) where its maximum value is $0.5c$.}
		\label{VELOCITY2}
\end{figure}

We remark that $\bm{v}$ is the velocity of the outflow, and is not the same as the velocity of the field lines. The velocity of the field lines in units of $c$ is given by
\begin{eqnarray}
\bm{V}_{F}=\frac{\bm{E}\times \bm{B}}{B^{2}}=\frac{(-\bm{v}\times \bm{B})\times \bm{B}}{B^{2}}=\bm{v}-\frac{\bm{v}\cdot\bm{B}}{B^{2}}\bm{B}\,,
\end{eqnarray}
thus for $\bm{v}$ along the $z$ axis it is
\begin{eqnarray}
\bm{V}_{F}=-\frac{v_{z}B_{\phi}B_{z}}{B^{2}}\hat{\bm{\phi}}+\frac{v_{z}B_{\phi}^{2}}{B^{2}}\hat{\bm{z}}\,.
\end{eqnarray}
There is a difference between $\bm{v}$ and $\bm{V}_{F}$ by the projection of $\bm{v}$ on $\bm{B}$. $\bm{V}_{F}$ has both an axial component and a toroidal component, Figs~(\ref{VELOCITY1}), (\ref{VELOCITY2}). Note that the plasma can drift along the field lines, thus it is not necessary that it moves with the $\bm{V}_{F}$ velocity, but in principle it may move with any velocity which is the sum of $\bm{V}_{F}$ and a component parallel to the magnetic field. The addition of a velocity component parallel to the magnetic field  has no effect in the dynamics of the problem in the absence of inertia or for an appropriate drift along the field lines so that the particles perform no azimuthal motion \citep{Gourgouliatos_Lynden-Bell:2011}. When inertia is taken into account the dynamics of the problem change as the first term in equation (\ref{MOM}) gives a centrifugal force when there is an azimuthal component in the velocity. We consider this force in the simulations below. Note also that in the demand of a smooth transition from the jet to the external medium $v_{z}$ needs to go to zero at the boundary of the cylinder, otherwise there will be a discontinuity in velocities in this layer. 
\section{MHD simulations of jet propagation}
\label{sims}
One of the main advantages of the magnetic field distribution proposed in this paper is that it provides a transition through the jet boundary  $R_{\rm jet} \equiv 1$ where no surface force is present.

Simulations of jet propagation for helical magnetic fields often rely on a toroidal magnetic field distribution which introduces spurious relaxation processes along the jet boundary. The problem lies in the fact that one usually injects a magnetized jet with a helical field into an ambient gas initially without any toroidal field. Examples of this approach are the seminal papers by \citet{clar86} and \citet{lind89}, 
but also many follow up simulations \citep{koes90, fran98, ston00, onei05}, and more recently the relativistic cases presented by \cite{kepp08, mign10}. Other authors have applied a force-free field distribution over the whole computational domain \citep{todo92,todo93}, i.e. a toroidal
field also in the ambient gas.

For example, it is possible to choose a sinusoidal function which leads to a vanishing toroidal magnetic field
at the jet boundary, but this is not a force-free field configuration. Thus, Lorentz forces within the jet will distort the injected
jet material. Furthermore, applying a vanishing toroidal field outside the jet (for $R> R_{\rm jet}$), the derivative $\partial B_{\phi} / \partial R$ becomes infinite and thus the Lorentz force at this radius. Another option is to choose a linearly increasing toroidal field, which is force-free across the jet and maintains the injected jet structure, however, since it is set to zero for $R > R_{\rm jet}$, it introduces again an infinite Lorentz force.

A more elaborate initial and boundary condition has been applied by \citet{kepp08}, who solved the full MHD equations 
to find a jet inlet in radial force-balance. However, even in this case the toroidal field distribution is cut off at the jet radius, potentially leading to relaxation forces along the jet boundary.

In this section we test the behaviour of the magnetic field configuration derived earlier in this paper in simulations. We have therefore performed axisymmetric relativistic MHD simulations of jet propagation applying the PLUTO 3.01 code \citep{mign07}.
We inject a jet with the velocity, pressure, and magnetic field profiles as derived in Section {\ref{GENERAL} into an ambient corona of vanishing magnetic field and uniform pressure. In addition to the properties derived, we need to define two
other dynamical parameters which are the jet density and the density of the ambient gas, $\rho_{\rm jet},\rho_{\rm ext}$. We do this by applying a polytropic equation of state, thus obtaining 
$\rho_{\rm jet} (R) = p_{\rm jet}(R)^{1/\gamma}/K_{\rm jet}$,
and 
$\rho_{\rm ext} (R) = p_{\rm ext}(R)^{1/\gamma}/K_{\rm ext}$.
Note that depending on the choice of $K_{\rm jet}, K_{\rm ext}$ there will be an entropy jump between the jet and the external 
material. Essentially, we inject jets with internal Alfv\'en Mach numbers of $2-10$ and fast magnetosonic Mach numbers of $>1$, as expected from MHD disk jets. Clearly, the hydrodynamical parameters will play an essential role in the stability characteristics of these jets. However, a full stability analysis is beyond the scope of this paper as it would require also a 3D-treatment. Here we restrict ourselves to a preliminary study of the morphological stability of the jet-ambient medium interface. For our simulations we apply a numerical grid of physical size  $(5\times 20)$ with a resolution of 100 equidistant cells between  $r=0$ and $r=1.5$ and 200 scaled cells between $r=1.5$ and $r=5$.  In $z$-direction we have 100 equidistant cells till $z=3$ and 250  scaled cells till $z=20$. We also have performed comparison  simulations with double resolution which did not show any significant differences.

The parameters of our simulation runs are listed in 
Tab.~\ref{tab:parajets}.
For comparison we provide as derived parameters the maximum
poloidal Alfv\'en Mach number 
$$ M_{\rm A} = \sqrt{ \frac{4\pi\rho_{\rm jet} u_{\rm z}^2}{B_{\rm z}^2} },$$ 
and the maximum poloidal fast ``cold" magnetosonic Mach number 
$$ M_{\rm FM} = \sqrt{ \frac{4\pi\rho_{\rm jet} u_{\rm z}^2}{B_{\rm z}^2+B_{\phi}^2} },$$
with the relativistic velocity defined as $u \equiv \Gamma v$ and
the Lorentz factor $\Gamma$.

To demonstrate the overall force-balance across the jet boundary, we plot the radial profile of radial Lorentz force component in
Fig.~\ref{fig:s07_lorentz}. Inside the jet the Lorentz force is non- vanishing, however it is balanced by the jet kinematic forces.
Towards the jet boundary the radial Lorentz force vanishes, and remains zero outside the jet.
\begin{figure}
\centering
\includegraphics[width=8cm]{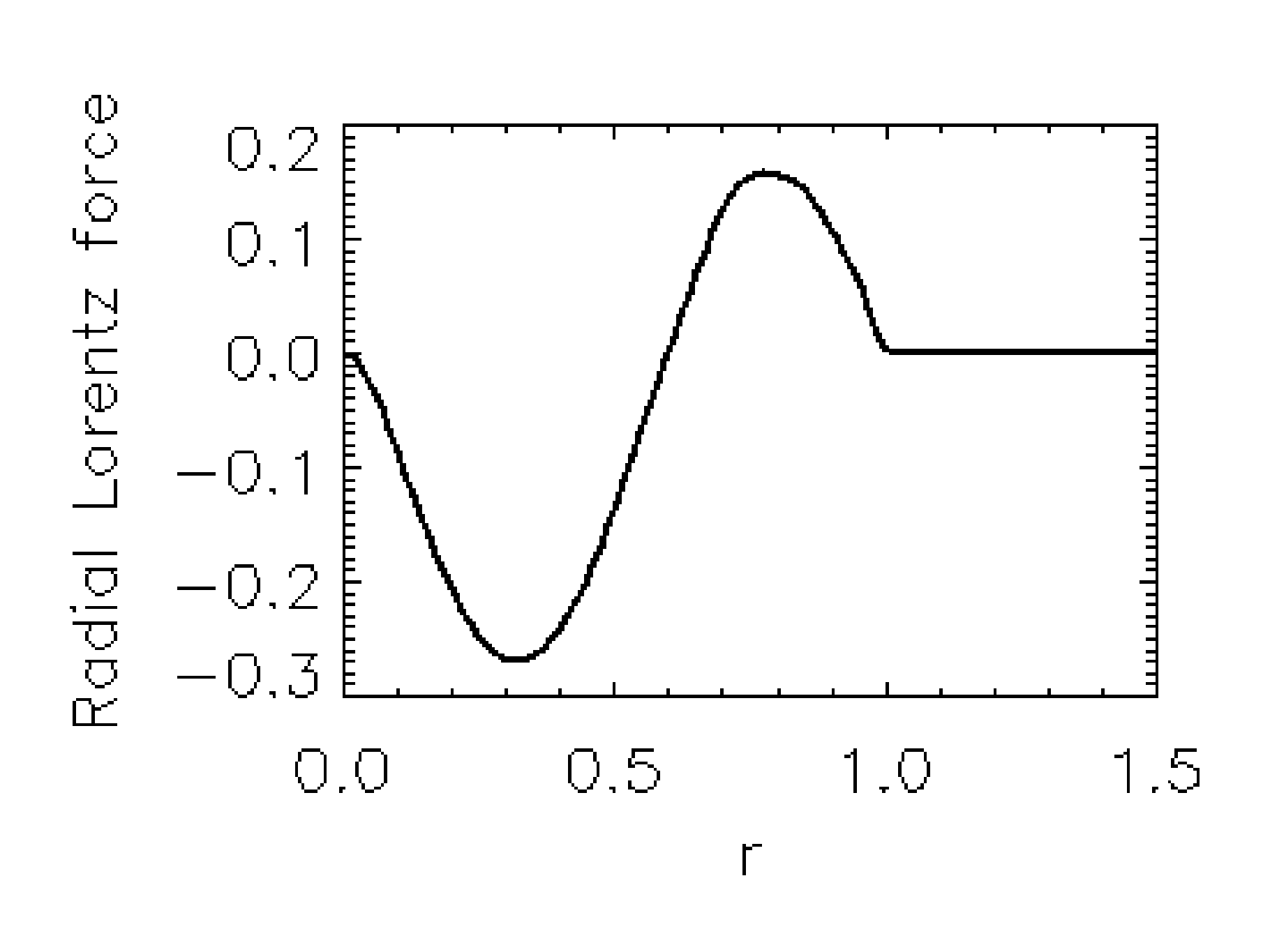}
\caption{Radial profile of Lorentz force radial component for 
simulation S07 after 500 dynamical time steps,
indicating a smooth transition from jet to ambient medium.
Shown is the profile along the first active zones.
\label{fig:s07_lorentz}
}
\end{figure}

Figure \ref{fig:sim_s06_dynam} shows the dynamical parameters for our reference simulation S06 at 500 dynamical time steps. 
The density and velocity distributions clearly demonstrate the smooth transition across the jet boundary, thanks to the  new analytical MHD solution introduced above. Along the axis reflection shocks develop, which, however, do not strongly affect the jet/ambient gas interface.

Figure \ref{fig:sim_mach} shows the poloidal Alfv\'en Mach numbers for simulations S04, S05, and S07.
This figure displays a smaller subset of the computational domain.
The flows are all well super Alfv\'enic, while only a small amount of the entrained material along the flow is sub Alfv\'enic.

The appearance of reflection shocks is a well-known feature in jet
propagation simulations
(see e.g. \citet{hard88, hard89, hard92,bodo04}).
Following \citet{hard88, hard89} we estimate the wavelength $\lambda$
of the reflections as
\begin{equation}
\lambda \simeq \frac{2}{m+1/2} \frac{M_{\rm j,int}}{1 + \sqrt{\eta}} R
\end{equation}
with the jet internal Mach number $M_{\rm j,int}$, the density
contrast $\eta = \rho_{\rm jet}/\rho_{\rm ext}$, the wave mode
number $m$, and half-thickness radius $R$.
With the usual definition of the generalized relativistic Mach number
$M \equiv \Gamma_{\rm jet} v / \Gamma_{\rm s} v_{\rm s}$
with the Lorentz factors for the jet bulk speed
$\Gamma_{\rm jet} \equiv (1-v_{\rm jet}^2)^{-1/2}$
and the sound speed
$\Gamma_{\rm s} \equiv (1-v_{\rm s}^2)^{-1/2}$,
and $v_{\rm s}$ is the sound speed normalized to the speed of light
$v_{\rm s} \equiv a/c$,
and with the sound speed
\begin{equation}
a \equiv \left[\frac{\gamma P}{\rho +  \gamma P /(\gamma-1)}\right]^{1/2}
\end{equation}
\citep{hard01}
For simulation run S07 we find $\eta = 0.5$, $P_{\rm jet} \simeq 1.48$,
a sound speed $a \simeq 0.19$, and with the Lorentz factors
$\Gamma_{\rm s} \simeq 1.02$, $\Gamma_{\rm jet} \simeq 1.25$,
resulting in a generalized Mach number $M \simeq 3.7$, and, thus,
a zeroth mode wave length $\lambda \simeq 7 R$.
If we take $R=0.5$, the radius where we measured the above-mentioned
quantities, the predicted wavelength is $\lambda \simeq 7 R$,
which is similar to our simulations result (see Fig.~10).

Figure \ref{fig:sim_s04_time-evol} demonstrates how the jet penetrates the ambient medium for simulation S04, showing the log-scale density
distribution of the whole computational domain at time 200 and 500, respectively.
This is an over dense jet with high (maximum) Lorentz factor 
$\Gamma = 7.1$ (i.e. $v_{\rm max} = 0.99$).
Again we see a smooth transition between jet and ambient gas
with only little entrainment going on.
In the case of this powerful jet, the wavelength of the reflection
shocks is longer.

\begin{table}
\begin{center}
\caption{Relativistic MHD simulations of jet propagation, applying model 
Fig.5.
Parameters common for all simulations are
$ F_p = -1.54 $,
$ \alpha_p =  5.14$,
$ c_p =  0.172$,
$ F_t = -1.10 $,
$ \alpha_t =  3.83 $,
$ c_t =  0.186 $.
Parameters specific for each simulation are listed below.
$p_{\rm ext} \equiv p_{p,0}$ or $p_{\rm ext} \equiv p_{t,0}$, 
respectively.
$v_{\rm jet} \equiv v_{z,0}$.
$K_{\rm ext} \equiv \rho_{\rm ext}^{-1}$.
$K_{\rm jet} \equiv \rho_{\rm jet}^{-1}$.
For comparison we provide as derived parameters the
internal Alfv\'en Mach number $M_{\rm A}$, and
internal fast magnetosonic Mach number $M_{\rm FM}$,
given as a typical value along the jet (naturally these parameters
vary substantially across the jet according to the velocity and
magnetic field profiles, see Fig.~10).
}
\label{tab:parajets}
\begin{tabular}{ccccc}
\noalign{\medskip}
\hline
\noalign{\smallskip}
                & Sim04    & Sim05  & Sim06 & Sim07   \\

\noalign{\smallskip}
\hline
\noalign{\smallskip}
$v_{\rm jet}$          & 0.999  & 0.5 & 0.5 & 0.9   \\
\noalign{\smallskip}
$\rho_{\rm jet}$       & 20.0 & 20.0 & 20.0 & 20.0  \\
\noalign{\smallskip}
$\rho_{\rm ext}$       & 5.0 & 100.0 & 100.0 & 100.0  \\
\noalign{\smallskip}
$p_{\rm ext}$          & 0.3 & 0.5 & 1.5  & 1.5      \\
\noalign{\smallskip}
\hline
\noalign{\smallskip}
$M_{\rm A}$           & $\simeq 5$ & $\simeq 2.5$ & $\simeq 3$ &  $\simeq 
9$  \\
\noalign{\smallskip}
$M_{\rm FM}$          & $\simeq 3$ & $\simeq 1.5$ & $\simeq 2$ &  $\simeq 
5$  \\
\noalign{\smallskip}
\hline
\end{tabular}
\end{center}
\end{table}
\begin{figure*}
\centering
\includegraphics[width=8.0cm]{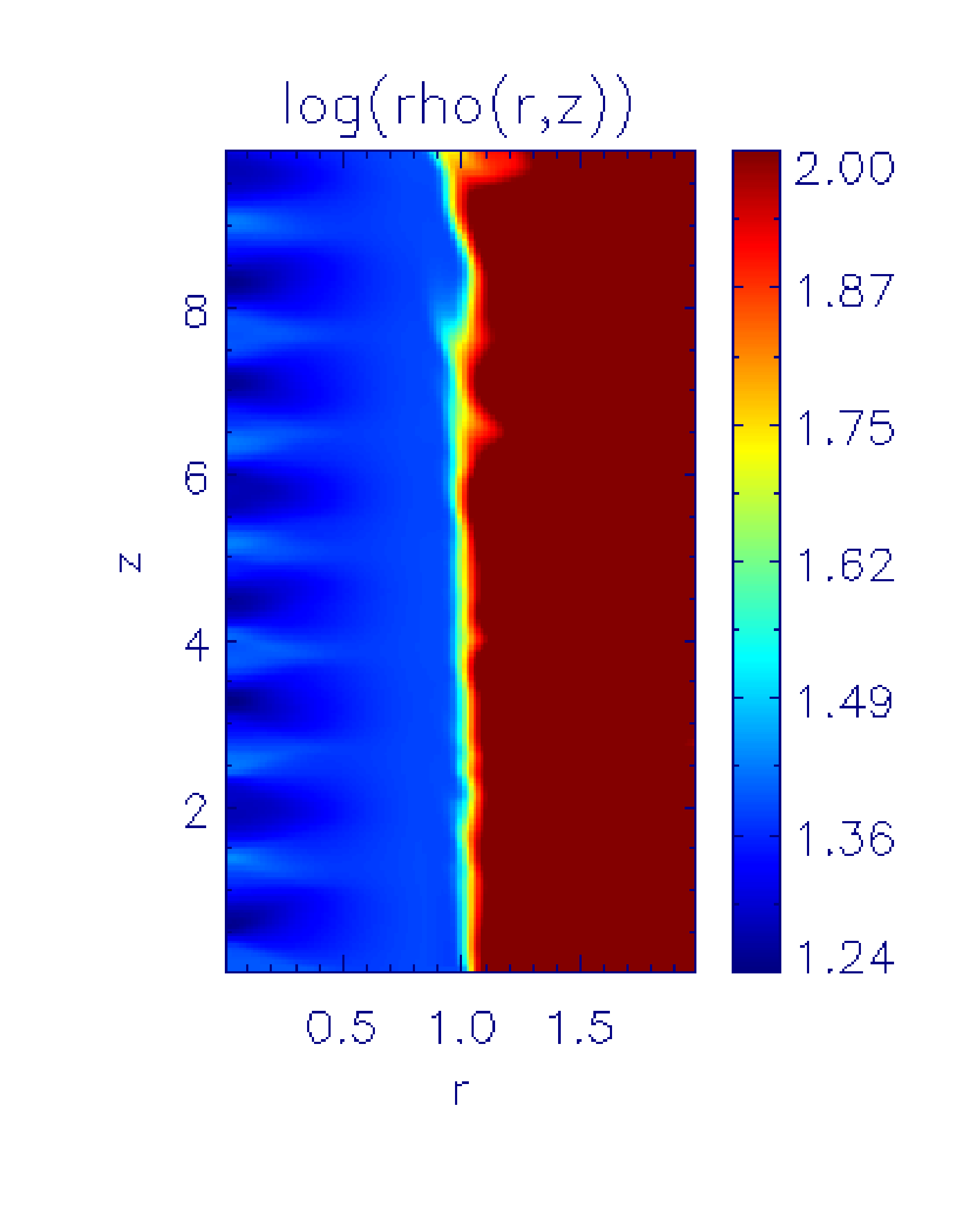}
\includegraphics[width=8.0cm]{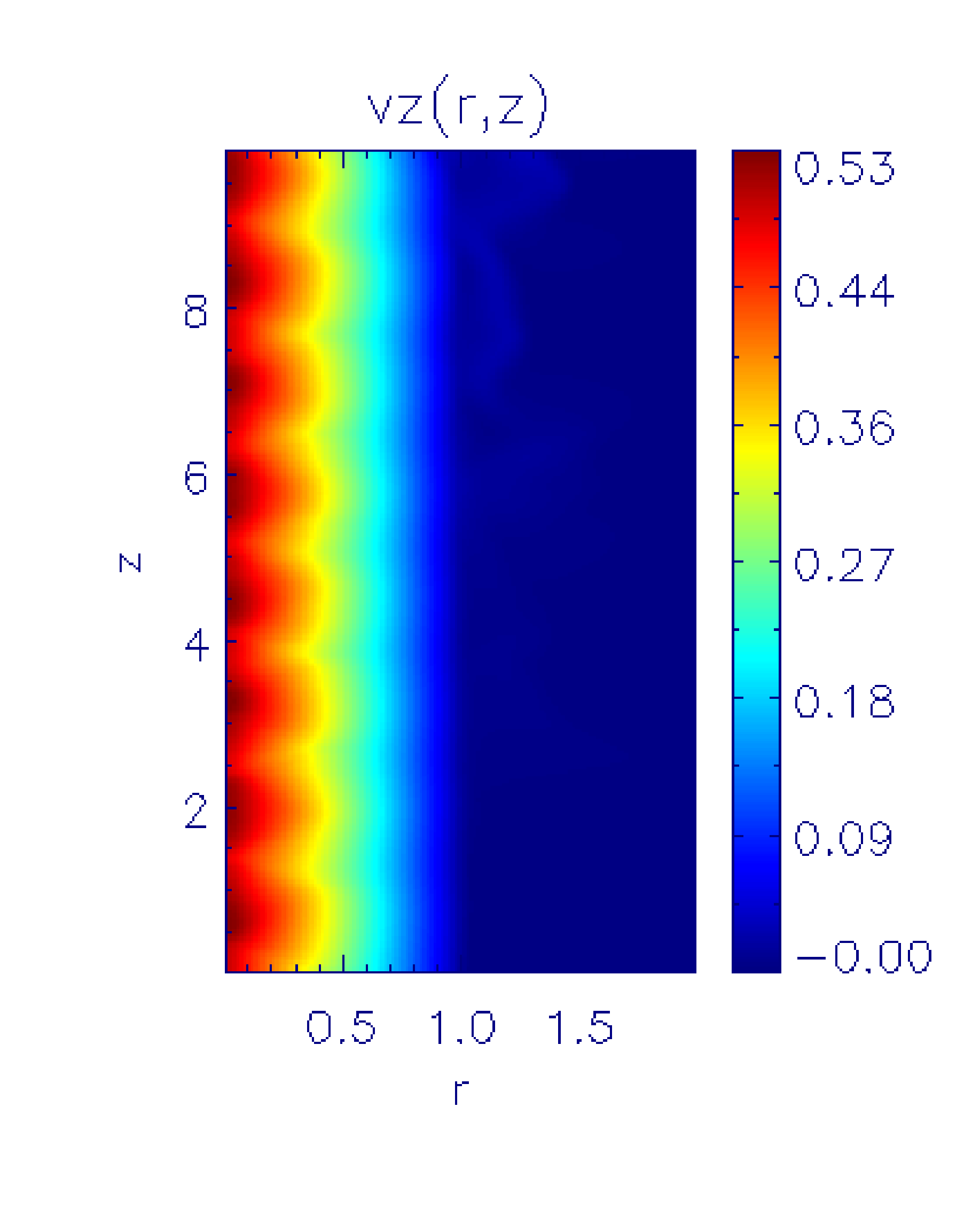}
\includegraphics[width=8.0cm]{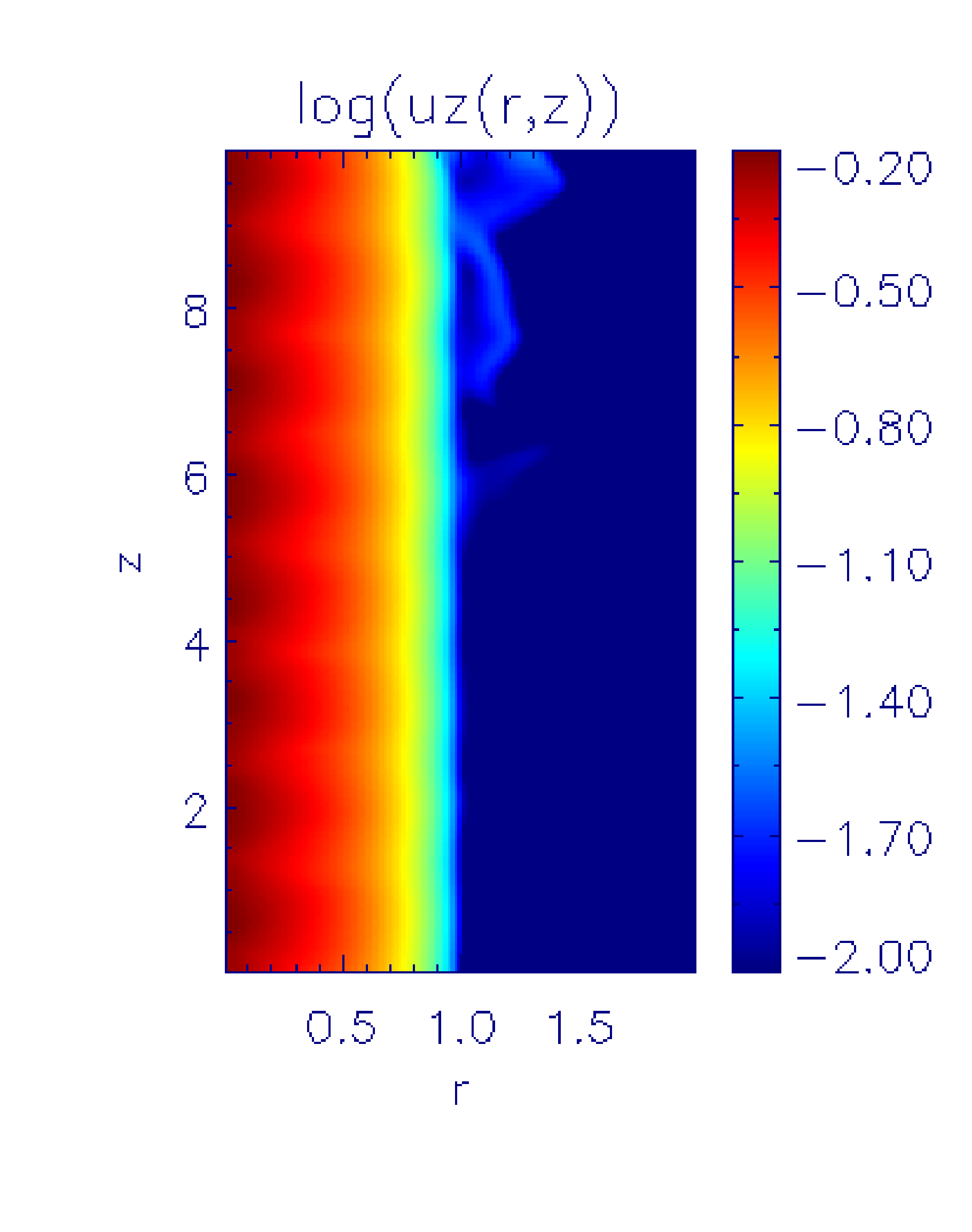}
\caption{Axisymmetric jet propagation simulation S06.
Density $\rho$, axial velocity $v_z$, 
and relativistic axial velocity $u_z = \Gamma v_z$
distribution after 500 dynamical time scales.
Note the size of the figure which is a $2 \times 10$ subset of
the whole $5 \times 30$ computational domain.
\label{fig:sim_s06_dynam}
}
\end{figure*}
\begin{figure*}
\centering
\includegraphics[width=8cm]{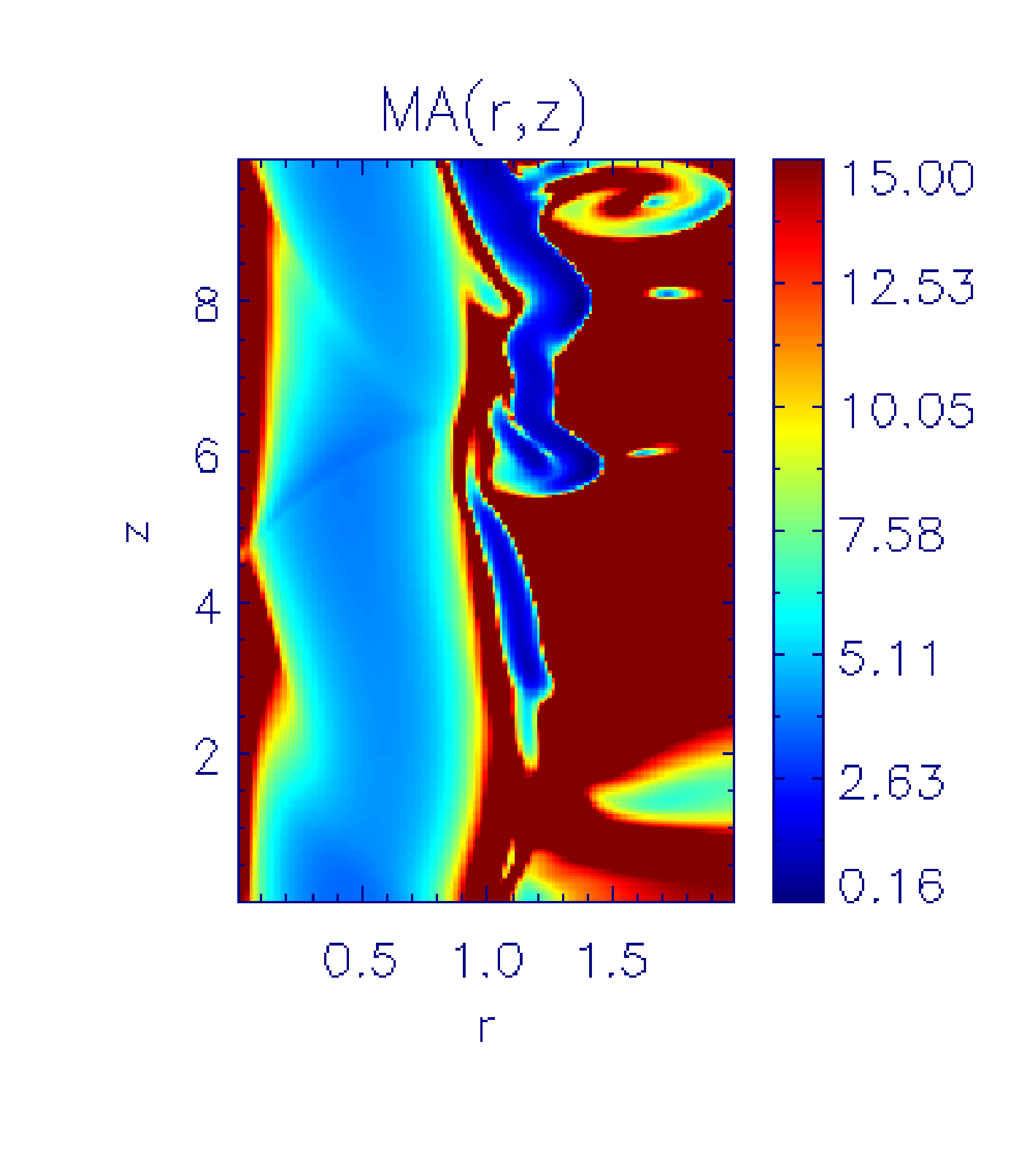}
\includegraphics[width=8cm]{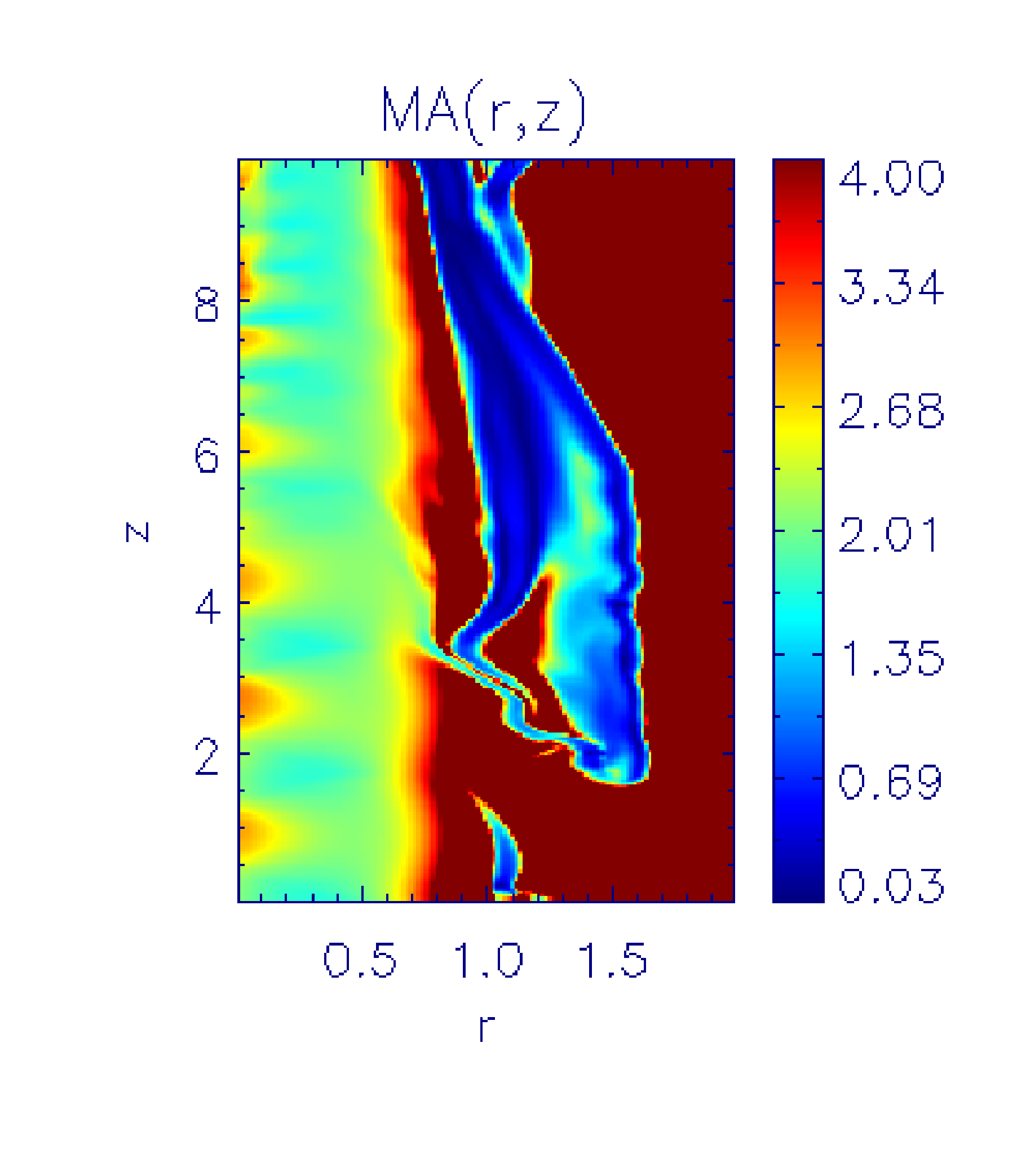}

\includegraphics[width=8cm]{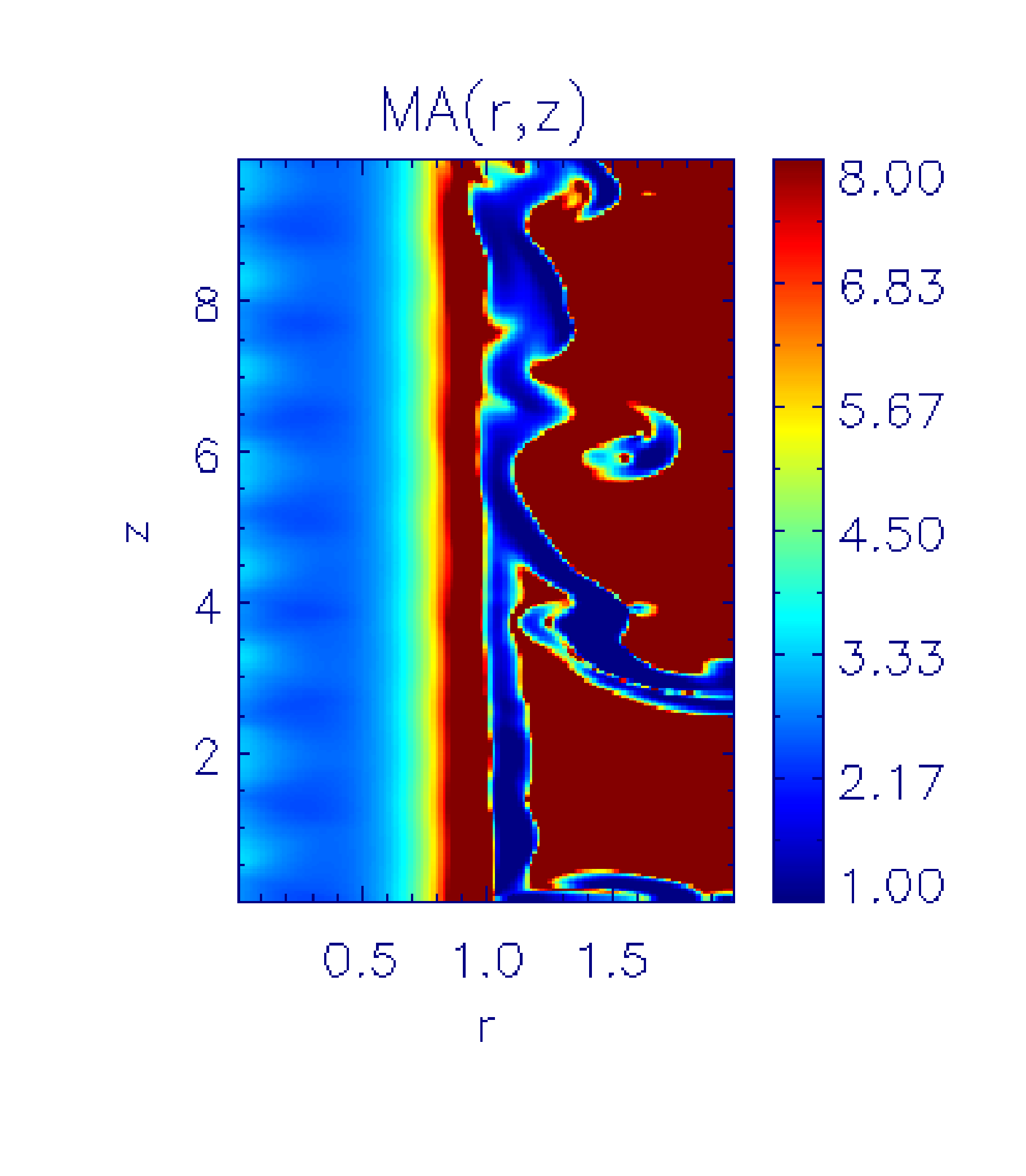}
\includegraphics[width=8cm]{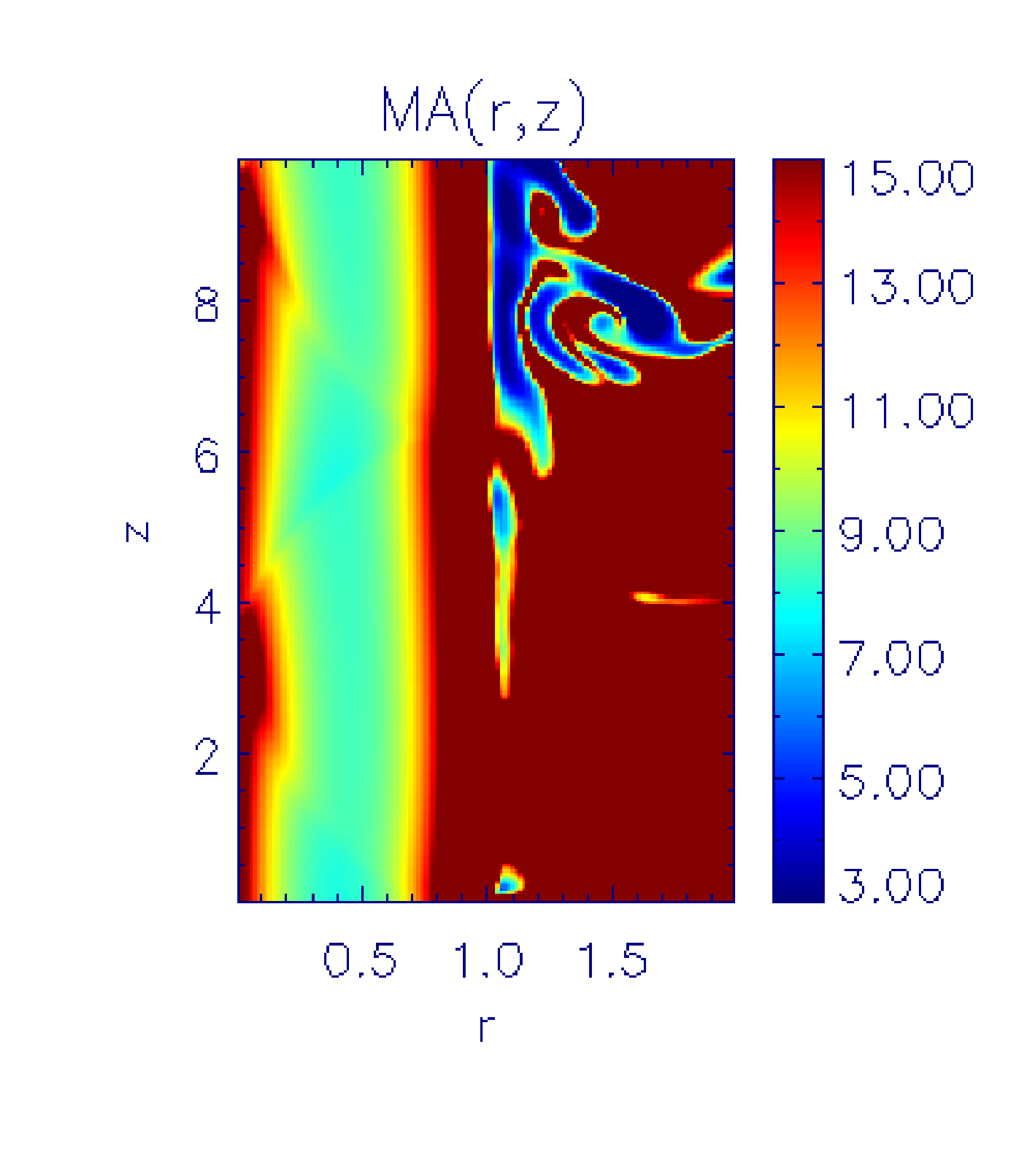}
\caption{Axisymmetric jet propagation simulations S04, S05, S06, S07.
 Shown is the poloidal Alfv\'en Mach number $M_{\rm A}(r,z)$.
 Note the (real) rather long wavelength of the reflections, 
 as only a subset of the whole computational domain is shown.
\label{fig:sim_mach}
}
\end{figure*}
\begin{figure}
\centering
\includegraphics[width=8cm]{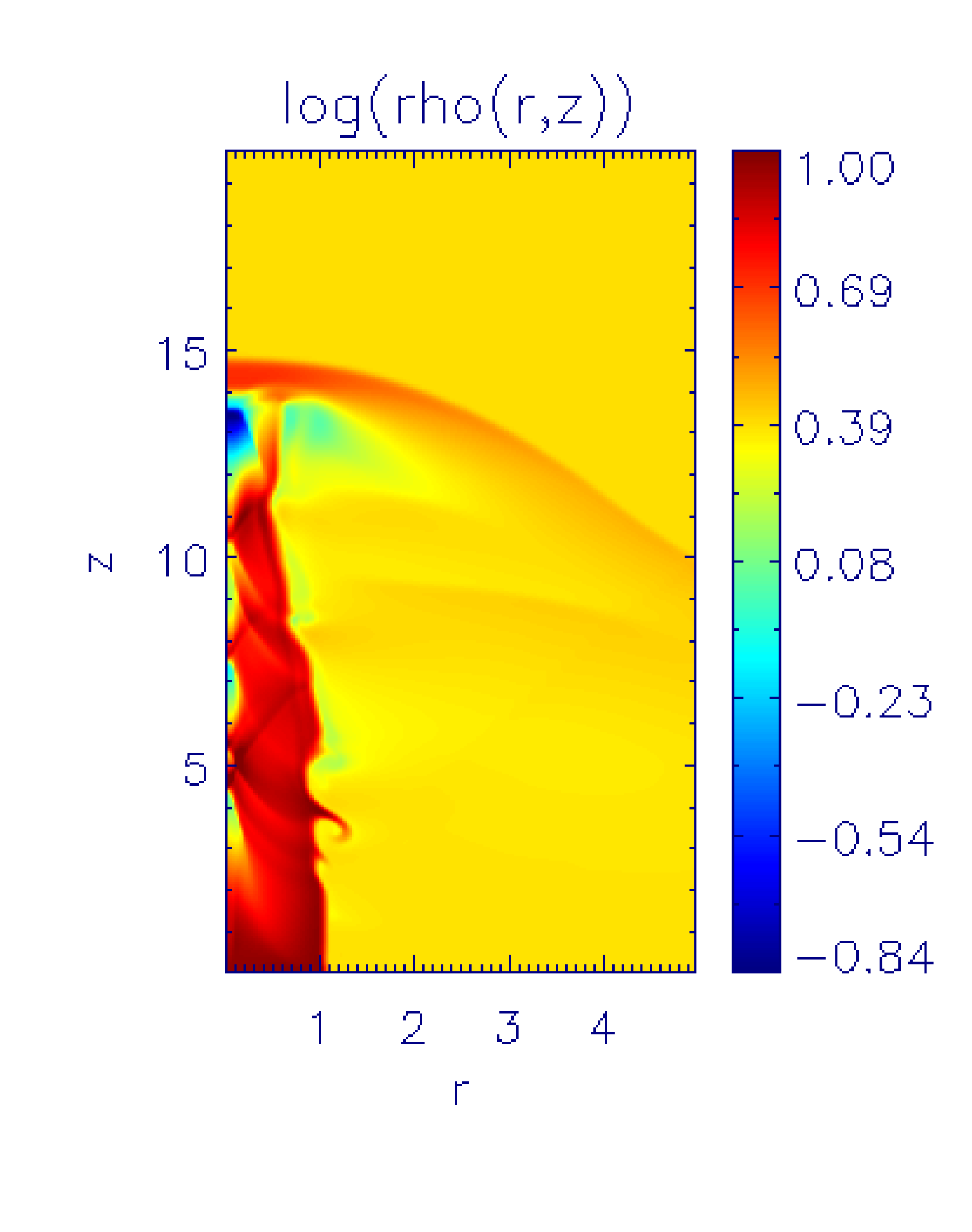}
\includegraphics[width=8cm]{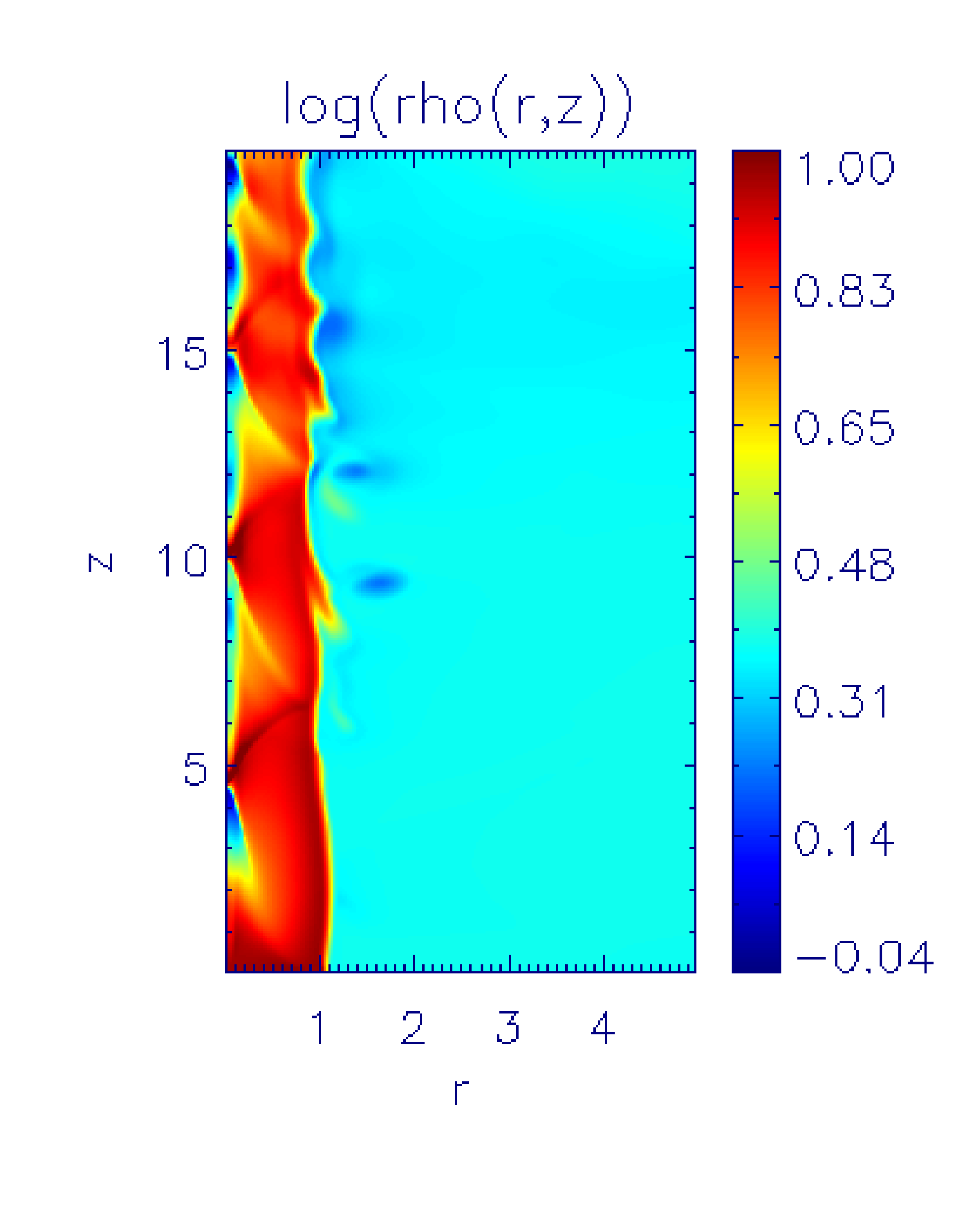}
\caption{Axisymmetric jet propagation simulation S04.
Large-scale density distribution after 200, 500 dynamical time scales,
respectively.
\label{fig:sim_s04_time-evol}
}
\end{figure}

\section{MHD stability}
\label{stability}
The stability of MHD equilibria has received much attention in the plasma confinement literature due to the violent instabilities to which Tokamak devices are prone \citep[see e.g.][]{Kadomtsev:1966,Bateman:1978,Freidberg:1987}.  A common technique for evaluating stability involves introducing perturbations of the form $\propto \exp{i(m\phi+kz-\omega t)}$ on top of the equilibrium configuration to examine how this affects the total MHD energy.  Analyses of this type have produced two well-known analytical criteria that provide simple tests for stability, the Kruskal-Shafranov (KS) criterion and the Suydam criterion.  

The KS criterion is a test for the current-driven $|m| =1$ mode instability.  This so called ``kink" mode is one of the most dangerous instabilities for magnetically dominated astrophysical jets because, unlike other modes, it displaces the fluid center of mass by distorting the jet cylinder into a helix-like configuration, thereby converting magnetic energy into kinetic energy.  The internal and external versions of this instability have already been the object of a number of studies in the astrophysical literature \cite[e.g.][]{Istomin:1996,Begelman:1998,Nakamura:2004,Giannios:2006,Narayan:2009}.  The KS criterion states that the jet is kink stable if
\begin{eqnarray}
\frac{2\pi R}{L_j}\left|\frac{B_z}{B_{\phi}}\right|>1,
\label{KS}
\end{eqnarray}
where $L_j$ is the length of the jet.  \cite{Narayan:2009} have found that this criterion is similar in form for force-free relativistic jets as long as one uses the jet frame magnetic fields in the above expression. Since realistic jets are conically expanding, ideal MHD suggests that $B_z'/B_{\phi} \ll 1$, rendering the kink mode particularly dangerous to jet stability.  However, we note that a dynamically significant amount of plasma pressure makes the jet less susceptible to the kink instability.

The Suydam criterion concerns the specific interplay between the curvature of the magnetic field lines and the pressure gradient.  For the stability of these local ``interchange" modes, it is necessary (but not sufficient) for the following equality to be satisfied for the entire jet profile:
\begin{eqnarray}
\frac{RB_z^2}{4\pi}\left(\frac{q'}{q}\right)^2+8p'>0\,,
\label{Suydam}
\end{eqnarray}
where $q=\frac{2 \pi R B_{z}}{L_{j}B_{\phi}}$.
Equation (\ref{Suydam}) shows that magnetic configurations with high enough values of $q'^2$ can overcome the destabilizing effects of a negative pressure gradient (i.e. $p'<0$).  Evaluating equation (\ref{Suydam}) for both equilibria reveals that the poloidal equilibrium is unstable between $R=0$ to 0.4, and the toroidal equilibrium is stable for all $R$.

As suggested by the stability of our solutions in the MHD propagation simulations (see \S\ref{sims}), the results from the KS and Suydam criterion should be approached with some skepticism.  The kink mode, mostly discussed in the AGN jet literature in the context of magnetically dominated jets, is not as important in jets where the plasma pressure is dynamically significant, as is the case with our solution where the plasma $\beta$ ($=8\pi p/B^2$) is of order unity in the outer region of the jet.  In the case of magnetically dominated jets, simple analytical kink mode analyses which suggest they are kink unstable \citep[e.g.][]{Begelman:1998} do not take into account complicating factors such as causality in conically expanding jets, gradual velocity shear, and field line rotation.  This may explain why some analytical stability analyses suggest jets are kink unstable while, in contrast, three-dimensional numerical simulations which do take into these complicating factors can produce jets which are kink stable \citep{McKinney:2009}.  In the case of pressure driven instabilities, the poloidal equilibrium is unstable per a naive application of the Suydam criterion.  Unfortunately, this criterion does not take into account bulk velocity and therefore has limited applicability for jets.  In the fusion confinement literature, \cite{Bondeson:1987} have extended the Suydam analysis to include velocity and have shown that the stability of localized pressure-driven modes depends on the quantity $M=v_z'\rho^{1/2}(B_zq'/q)^{-1}$, a form of Alfv\'enic Mach number.  If $M<\beta$ then velocity shear destabilizes local resonant modes; if $M>\beta$ such modes are stabilized, though other global modes are excited with slow growth rates \citep[for a discussion of pressure-driven modes in the context of jets, see][]{Longaretti:2008}.  Thus, the apparent stability of the poloidal equilibrium in our jet propagation simulations may be due to high velocity shear, though \cite{Bondeson:1987}'s work would need to be extended to the relativistic regime for a more thorough understanding of how velocity shear affects pressure-driven instabilities in relativistic jets.

\section{Synchrotron Jet Predictions}

The polarization and anisotropy of the resulting synchrotron emission from magnetic jets containing relativistic plasmas, allows high resolution telescopes to probe the structure of jet magnetic fields.  To do this, the transverse structure of the jet must be resolved, a feat that only very long baseline interferometric (VLBI) radio telescopes can perform on some active galactic nuclei jets.  The emission patterns of a transverse (compared to the jet symmetry axis) cut on a jet threaded with a large-scale helical magnetic field has already been studied using analytical models \citep[e.g.][]{Laing:1980,Lyutikov:2005,Clausen-Brown:2011} and numerical simulations \citep[e.g.][]{Broderick:2010, Porth:2011}. Following \cite{Clausen-Brown:2011}, we create polarization profiles by calculating the Stokes $Q$ and $I$ ($U=0$ since our chosen reference position angle is the jet direction projected onto the sky).  We assume the entire jet cylinder is filled with emitting electrons with a power-law electron distribution function, $dn=K_e E^{-p}dE$, where the density of particles does not vary with $R$.  Retaining the sign of $Q$ in calculating $\Pi$ gives the synchrotron radiation's electric vector position angle (EVPA) as well: when $Q$ is positive the EVPA is along the projected jet direction and when $Q$ is negative the EVPA is perpendicular to the projected jet direction.  Faraday rotation measure (RM) profiles are calculated as well using a Faraday rotating electron density that is constant with radius.

Figure \ref{pol} shows the calculated polarization and RM profiles using our solutions (labeled ``Poloidal" and``Toroidal") and the Lundquist field with two different cutoffs chosen (``Force-free I" and ``Force-free II") so they most resemble our solutions.  The similarity between our solutions and their force-free counterparts will make distinguishing them using VLBI profile difficult, especially considering that most AGN jets are barely resolved if at all. 

\begin{figure}
	\centering
		\includegraphics[width=0.48\textwidth]{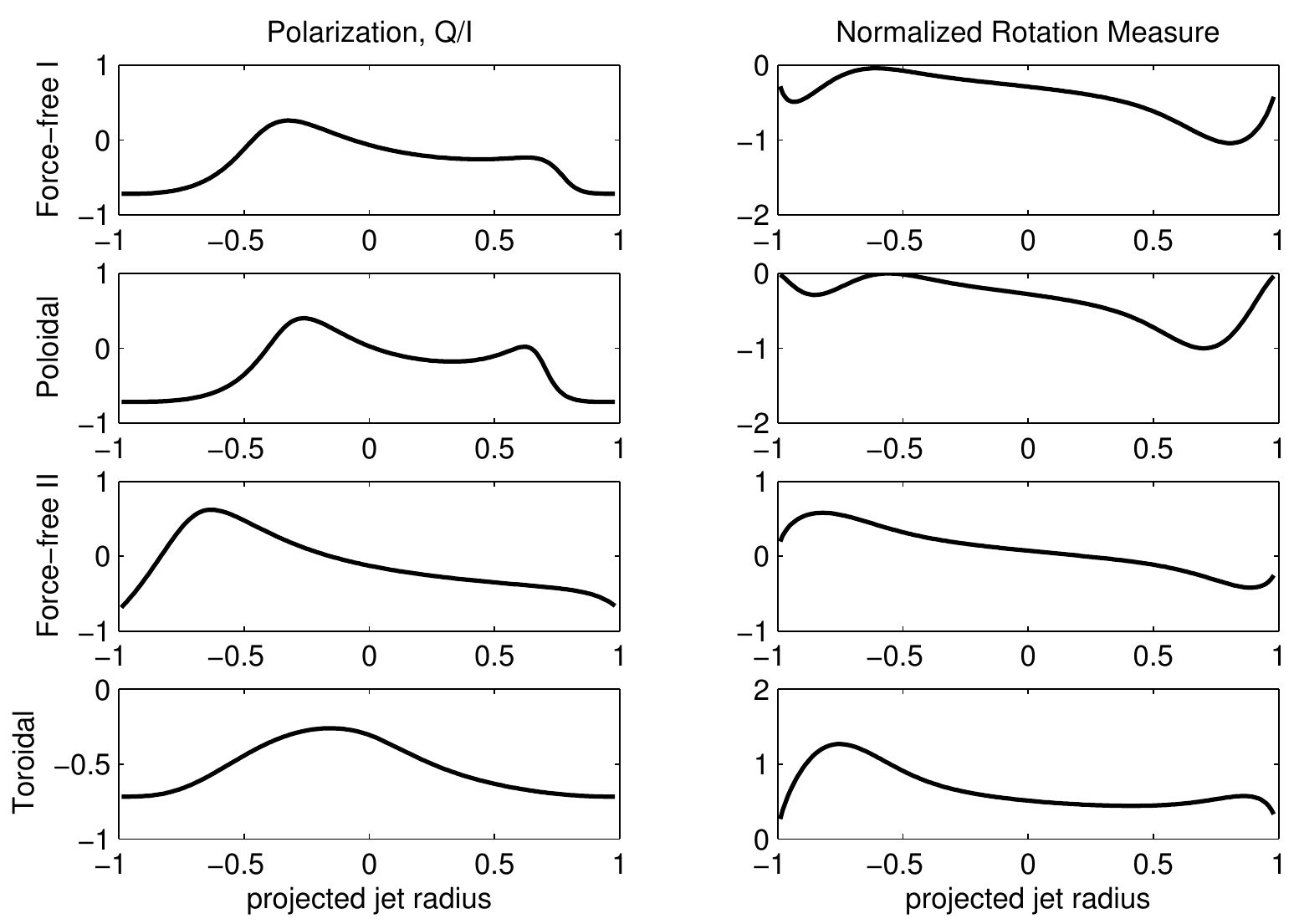}
		\caption{In the left column are plots of the fractional polarization of the jets as a function of projected radius for jet viewing angles of $\theta_{ob}=1/(2\Gamma)$.  The sign of the polarization refers to whether the associated EVPA is parallel ($Q/I>0$) or perpendicular ($Q/I<0$) to the jet axis.  The right column are plots of the Faraday RM that are normalized by the maximum change ($\Delta RM=RM_{max}-RM_{min}$) in RM.  Each of the four rows of plots refers to (a) the Lundquist solution where $B_{\phi}(R=1)=0$, (b) the $B_p$ equilibrium, (c) the  Lundquist solution where $B_z(R=1)=0$, and (d) the $B_t$ equilibrium.  Notice that the magnetic field structure for solutions (a) and (b) are similar in that they undergo a $B_z$ reversal, and solutions (c) and (d) are similar in that no magnetic field reversals take place.  It is for this reason that the plots on the first two rows resemble each other and the third and fourth row plots resemble each other.}
\label{pol}
\end{figure}

\section{Discussion}

In this work we have considered the magnetic field structure of a magnetized cylindrically symmetric jet. We have found analytical solutions for the field and the plasma pressure that is confined by an external medium without a singular current sheet, but through a smooth transition. The configuration satisfies the Grad-Shafranov equation subject to given boundary conditions, namely both the magnetic flux and its derivative have to be zero. Our assumption of cylindrical symmetry is valid in a jet's asymptotic region well downstream of the collimation region where the structure transitions from a broad outflow into a conical one with a small opening angle. Indeed, a jet powered by a magnetic mechanism \citep{Blandford:1977, Blandford:1982, Lynden-Bell:2003, Uzdensky:2006} shall start with a weakly collimated conical shape and, further downstream, become collimated \citep{Vlahakis:2003,Komissarov:2007, Porth:2010}. This is supported by observations of highly collimated jets observed in broad variety of astrophysical structures ranging from Herbig-Haro objects to AGN jets \citep{Livio:1999,Reipurth:2001}. M87 contains a well studied example of a collimated jet in which \cite{Junor:1999} have observed a broad structure very close to the origin with a half opening angle of $60^{\circ}$; collimation starts at around 30--100 Schwarzschild radii from the black hole and continues out to 1000s of Schwarzschild radii. The opening angle of the M87 jet at the parsec scale is around $6-7^{o}$ indicating  recollimation \citep{Kovalev:2007}. \cite{Pushkarev:2009} have found from a statistical study of relativistic jets that their intrinsic opening angles are around $2-3^{o}$. Obviously, a global jet model shall account for these variations, but the use of a cylindrical geometry locally for regions where the opening angle is small, is a reasonable approximation.  In our solution we have imposed a constant pressure environment which confines the jet to a cylinder. A more realistic model shall contain a varying pressure which decreases with distance from the origin and leads to jets of more complicated profiles, i.e.~conical or paraboloidal. From an other point of view, close to the launching region the physical processes do not simplify to force equilibrium as we have assumed in our model, making our solution applicable away from the launching region. 

Within the limits of our approach our solutions might be interesting in the light of the FR-I/FR-II dichotomy. FR-I sources have wider opening angles and entrainment with their environment and are therefore more unstable to Kelvin-Helmoltz modes \citep[e.g.][]{Bicknell:1995}, while FR-II sources are more collimated and have a smoother interface with their environment in a manner more consistent with our solutions.

Our solutions retain the choice of constant $\alpha$, usually applied in force-free structures, because it is known to give the most stable amongst the force-free fields. The linear relation of the pressure on the fluxes allows analytical solutions. This gives two classes of fields: one where the axial field is reversing and another where the axial field always points in the same direction. These solutions can be generalized to include relativistic outflows. We then use these solutions as starting points in simulations. Through these preliminary simulations we verify their stability and we also check the effects of centrifugal forces, as the particles do not move parallel to axis but follow the helical motion of the field lines, leading to a slightly different profile compared to the analytical solution. Finally the observational profiles predicted by this structure give reasonable results, but not clearly distinguishable from the ones predicted by force-free structures.  

We consider these solutions as a guide to investigate realistic configurations. They are simple enough and analytical so that they can be used to parametrize physical systems or applied as trial cases in numerical simulations. Given an appropriate choice of initial magnetic field, mass load and velocity profile a jet can be simulated. Possible applications are relativistic jets emanating from quasars and microquasars and in the non-relativistic context protostellar jets.

\section*{acknowledgements}
This research is supported by NASA grant NNX09AH37G. C.F. thanks Andrea Mignone and the PLUTO team for the possibility to 
apply their code. The authors are grateful to an anonymous referee for his insightful comments.

\bibliographystyle{mn2e}
\bibliography{BibTex}
\label{lastpage}

\end{document}